\documentclass[11pt,a4paper]{article}
\pdfoutput=1
\usepackage{jheppub1}
\usepackage{amsfonts,amsmath,amssymb}
\usepackage{enumerate}
\usepackage{hyperref}
\usepackage{graphicx}
\usepackage{caption}
\usepackage{subcaption}

\usepackage{enumitem}
    
\newcommand{\be}{\begin{equation}}
\newcommand{\ee}{\end{equation}}

\newcommand{\bea}{\begin{eqnarray}}
\newcommand{\eea}{\end{eqnarray}}

\newcommand{\bes}{\begin{subequations}}
\newcommand{\ees}{\end{subequations}}

\newcommand{\nn}{\nonumber}

\makeatletter
\newcommand{\citenosort}[1]{\begingroup\def\@addto@cite@list{\@cite@dump@now}\cite{#1}\endgroup}
\makeatother

\title{
\begin{center}
Uplifting dyonic AdS$_4$ black holes \\on seven-dimensional Sasaki-Einstein manifolds
\end{center}}

\author[a]{Hyojoong Kim}
\author[a, b]{and Nakwoo Kim}

\affiliation[a]{Department of Physics 
and Research Institute of Basic Science, \\ Kyung Hee University, 
Seoul 02447, Republic of Korea}
\affiliation[b]{School of Physics, Korea Institute for Advanced Study, Seoul 02445, Republic of Korea}

\emailAdd{h.kim@khu.ac.kr}
\emailAdd{nkim@khu.ac.kr}

\abstract{We revisit non-rotating, dyonically charged, and supersymmetric AdS$_4$ black holes, which are solutions of $\mathcal{N}=2$ gauged supergravity with vector- and hyper-multplets. Uplifting the near horizon solutions to $D=11$ supergravity on seven-dimensional Sasaki-Einstein manifolds, we show that dyonic AdS$_4$ black holes correspond to AdS$_2$ solutions with electric and magnetic baryonic fluxes in $D=11$ supergravity. 
We identify the \emph{off-shell} AdS$_4$ black hole solutions with parameters of $D=11$ AdS$_2$ solutions without imposing the equations of motion.
We calculate the entropy of dyonic black holes, carefully analyzing the Page charge quantization conditions.}

\begin{document}
\maketitle
\flushbottom
\section{Introduction}
The microscopic understanding of the Bekensteien-Hawking entropy of black holes has been a touchstone for any theory of  quantum gravity.
Recently, via AdS/CFT correspondence, the microscopic entropy was successfully computed for a class of supersymmetric and asymptotically AdS black holes.
The breakthrough was triggered through two concrete examples.
 The topologically twisted index of ABJM theory at large $N$ exactly matches the entropy of  static, magnetically charged AdS$_4$ black holes in dual supergravity \cite{Benini:2015eyy, Benini:2016rke}. 
 For the rotating, electrically charged AdS$_5$ black holes, on the other hand,  the entropy was reproduced by computing the superconformal index of ${\mathcal{N}}=4$ super Yang-Mills theory \cite{Cabo-Bizet:2018ehj, Choi:2018hmj, Benini:2018ywd}. 
See, for a review, \cite{Zaffaroni:2019dhb} and the references therein for the related studies.

As it is very well known, the entropy of a black hole is computed by the Bekenstein-Hawking formula
\be
S_{\textrm{BH}}= \dfrac{A}{4 G_N^{(4)}},
\ee
where $A$ is the area of the event horizon and $G_N^{(4)}$ is the four-dimensional Newton's constant. 
On the other hand, a new method to calculate the entropy was  recently proposed in \cite{Couzens:2018wnk} and discussed further in \cite{Hosseini:2019use, Gauntlett:2019roi, Hosseini:2019ddy, Kim:2019umc, Gauntlett:2019pqg, vanBeest:2020vlv}.
In \cite{Couzens:2018wnk}, the authors revisited AdS$_2 \times Y_9$ solutions in eleven-dimensional supergravity \cite{Kim:2006qu, Gauntlett:2007ts}. By imposing the supersymmetry conditions and suspending the imposition of the equations of motion, they constructed the so-called supersymmetric action $S_{\textrm{SUSY}}$. One can obtain the supersymmetric solutions by extremizing this nine-dimensional action $S_{\textrm{SUSY}}$. 
Then, the authors considered magnetically charged AdS$_4$ black holes \cite{Caldarelli:1998hg} in minimal gauged supergravity and 
uplifted the near horizon solutions to $D=11$ supergravity on an arbitrary seven-dimensional Sasaki-Einstein manifold SE$_7$ \cite{Gauntlett:2007ma}. Putting everything together, it was shown that $S_{\textrm{SUSY}}$ when extremized gives the entropy of this class of black holes.\footnote{For $D=10$ IIB supergravity, the story is parallel to M-theory case. The AdS$_3$ solutions in IIB supergravity \citenosort{Kim:2005ez, Gauntlett:2007ts}  were reformulated as variational problems in  \cite{Couzens:2018wnk} . One can holographically calculate the right moving central charge $c_{R}$ of two-dimensional dual field theories \cite{ Gauntlett:2018dpc, Hosseini:2019use}.}

This extremization principle provides a very powerful technique for calculating the black hole entropy
when $Y_9$ is a fibration of seven-dimensional Sasaki-Einstein manifold over a two-dimensional Riemann surface $\Sigma_g$ with genus $g$. 
Because all the information required to calculate $S_{\textrm{SUSY}}$ is just the topological data, i.e., the toric vectors of cone $C(SE_7)$, one can calculate the entropy without knowing the explicit form of the metric of black hole solutions.
It was first investigated in AdS$_3$ solutions of IIB supergravity in \cite{Gauntlett:2018dpc} and applied to $D=11$ supergravity for $S^7$ \cite{Hosseini:2019use} and for other SE$_7$ manifolds \cite{Gauntlett:2019roi, Hosseini:2019ddy, Kim:2019umc}.

 In this paper, we are interested in the AdS black holes in four-dimensional gauged supergravity obtained as consistent truncations of $D=11$ supergravity on SE$_7$. Then, the four-dimensional massless vector fields are classified into two groups. Some of the vector fields are originated from the isometries of SE$_7$ while the others, called Betti vector fields, are from the reductions of three-form potential $A_3$ on non-trivial two-cycles in SE$_7$. They are related to mesonic vs. baryonic flavor symmetries of dual boundary field theories, respectively.

First let us consider the simplest Sasaki-Einstein seven-manifold $S^7$. Since there is no non-trivial two-cycle in it, all the massless vector fields come only from the isometries of $S^7$. Then, the black holes have mesonic charges only. For example, it is well known that $D=11$ supergravity is consistently truncated to $D=4,\, \mathcal{N}=2,\,  U(1)^4$ gauged supergravity where $U(1)^4 \subset SO(8)$ are isometries of $S^7$ \cite{Duff:1999gh, Cvetic:1999xp}. This theory is called STU supergravity, where the black hole solutions with the spherical horizon were studied in \cite{Cacciatori:2009iz, DallAgata:2010ejj, Hristov:2010ri}. As we mentioned, the entropy of magnetically charged black hole is reproduced by computing the topologically twisted index  \cite{Benini:2015eyy} and also by solving the extremization problem \cite{Hosseini:2019use}.

Now let us consider topologically non-trivial SE$_7$ manifolds having non-trivial two-cycles. In this case, the Betti vector fields come into play in addition to the vector fields from isometries and the black holes have both mesonic and baryonic charges, in general.
However, to the best of our knowledge, neither black hole solutions nor the topologically twisted indices for general mesonic and baryonic charges has been computed.

On the one hand, explicit four-dimensional black hole solutions or AdS$_2$ solutions in M-theory with mesonic charges, except for that of graviphoton, are currently not available. 
 However, if we assume that such a black hole solution exists, then we can compute the entropy by solving variational problems and show perfect agreement with the field theory computations \cite{Gauntlett:2019roi, Hosseini:2019ddy, Kim:2019umc}.\footnote{For $\mathcal{N}=2$ Chern-Simons theories with chiral quiver, it is known that the matrix model to calculate the free energy on $S^3$ is not working \cite{Jafferis:2011zi}. An alternative prescription for calculating the free energy was proposed by \cite{Gulotta:2011si, Gulotta:2011aa} and further studied in \cite{Kim:2012vza, Kim:2019umc}. }
On the other hand, the explicit form of black holes with baryonic charges are well known.
The consistent truncations, which keep the Betti vector fields as well as the graviphoton, are constructed in \cite{Cassani:2012pj} and  the four-dimensional AdS black hole solutions with baryonic charges were studied in \cite{Halmagyi:2013sla}. Furthermore, AdS$_2$ solutions in M-theory \cite{Gauntlett:2006ns, Donos:2008ug} also have baryonic charges.
See also \cite{Klebanov:2010tj, Donos:2011pn, Donos:2012sy, Azzurli:2017kxo, Hong:2019wyi}. 
Given the explicit black hole solutions with baryonic charges, one can easily compute the Bekenstein-Hawking entropy, which can be successfully reproduced by solving variational problem \cite{Gauntlett:2019roi, Kim:2019umc}.
However, it is not known how to calculate the entropy on the dual field theory side because the topologically twisted index at large $N$
is independent of baryonic charges \cite{Hosseini:2016tor, Hosseini:2016ume}. 

Hence, the correspondence between the black hole entropy and the topologically twisted index does not hold for the black holes with both baryonic and mesonic charges,  because the entropy is known for black holes with baryonic charges only and the topologically twisted index depends on mesonic charges only. However, using the extremization principles, we can investigate the general cases with both baryonic and mesonic charges. Then, we expect that this method gives a guideline to resolve this puzzle and establish a precise correspondence.

Although the extremization principle provides a very powerful method to calculate the black hole entropy, it is applicable to a restricted class of black holes, i.e., four-dimensional magnetically charged black holes. One can think of various generalizations of this method to incorporate, for example, dyonically charged black holes, rotating black holes\footnote{Recently, $S_{\textrm{SUSY}}$ for rotating AdS$_4$ black holes is constructed in \cite{Couzens:2020jgx}.}, black holes in dimensions higher than four, etc. 
However, establishing the variational problems for these black holes from scratch is a daunting task and beyond the scope of this paper.

In this paper, instead of tackling the variational problem for dyonic black holes, we restrict ourselves to the explicit solutions of dyonic AdS$_4$ black holes.
For the black holes in STU supergravity, whose massless vector fields come from the isometries of $S^7$, the agreement between the entropy of the black hole and the topologically twisted index for ABJM theory also holds for the dyonic case \cite{Benini:2016rke}. However, the situation is more involved
for the black holes which we revisit in this paper. The dyonically charged AdS$_4$ black holes in \cite{Halmagyi:2013sla} were studied in four-dimensional $\mathcal{N}=2$ gauged supergravity, which is obtained by a consistent truncation of eleven-dimensional supergravity on a seven-dimensional homogenous manifold with $SU(3)$-structure \cite{Cassani:2012pj}. One of such manifolds we will focus on is $M^{111}$ where the truncation yields two massless vector fields: a graviphoton and one Betti vector filed. Therefore, the black holes we are interested in have baryonic charges and their dual field theoretic computation of the entropy at large $N$ is not known as we mentioned earlier.

Furthermore, while the STU black holes are known analytically, the black hole solutions we revisit in this paper can be studied numerically, which interpolate between AdS$_4$ and AdS$_2 \times \Sigma_g$. Only the near horizon solutions are known analytically \cite{Halmagyi:2013sla}. Hence, we can uplift this near horizon solutions and obtain AdS$_2$ solutions in eleven-dimensions. On the other hand, there are previously well-known AdS$_2$ solutions in $D=11$ supergravity discovered a decade ago \cite{Gauntlett:2006ns, Donos:2008ug}. These AdS$_2$ solutions,  which we review in the appendix \ref{ads2-review}, are classified into two groups according to the explicit form of four-form flux.\footnote{These solutions are revisited in \cite{Azzurli:2017kxo, Hong:2019wyi}. Recently, AdS$_2$ solutions outside this classification have been studied in \cite{Couzens:2020jgx, Ferrero:2020} as the near horizon solutions of the rotating AdS$_4$ black holes uplifted to $D=11$ supergravity. Furthermore, other AdS$_2$ solutions were discussed in type IIB supergravity \cite{Lozano:2020txg} and massive type IIA supergravity \cite{Lozano:2020sae}.
}
 
Though it was already mentioned in \cite{Halmagyi:2013sla} that uplifting dyonic AdS$_4$ black holes\footnote{Uplifting to $D=11$ was also discussed in \cite{Katmadas:2015ima}.} will be included in this class of AdS$_2$ solutions, we will explicitly study the uplifting and clarify the relations between them. 
In \cite{Kim:2019umc}, we already showed that the metric uplift of magnetically charged black hole solution corresponds to that of AdS$_2$ solution with electric four-form flux \cite{Gauntlett:2006ns, Azzurli:2017kxo}. In this paper, we complete the analysis by uplifting the flux as well. Then we move on to the dyonic black hole case. We identify the uplifted solution of dyonically charged black hole and AdS$_2$ solution with non-vanishing magnetic four-form flux \cite{Donos:2008ug, Azzurli:2017kxo}.
It implies that turning on the additional electric charges on the magnetically charged AdS$_4$ black holes corresponds to turning on the magnetic, internal four-form flux of $D=11$ supergravity. 

We also establish the precise dictionary between the four-dimensional quantities of AdS black holes \cite{Halmagyi:2013sla} and the parameters in AdS$_2$ solutions in M-theory \cite{Gauntlett:2006ns, Donos:2008ug}. 
These identifications lead to the following consequences. Firstly, it provides the expressions for the electric and magnetic charges of the four-dimensional black hole solutions in terms of the parameters, which describe the eleven-dimensional AdS$_2$ solutions.
Secondly, the identifications can be extended to \emph{off-shell}. For the parameters in AdS$_2$ solutions in M-theory, we should impose a constraint between them, which solves the equation of motion \cite{Gauntlett:2006ns, Donos:2008ug}. Here off-shell means that the identifications hold without imposing this constraint. In this sense, the four-dimensional black hole solutions written in terms of the parameters in the eleven-dimensional solutions are dubbed as off-shell solutions.
Thirdly, we calculate the off-shell expression for Bekenstein-Hawking entropy of the dyonically charged AdS$_4$ black holes. When we have the non-trivial internal four-form flux, we should consider the flux quantization with Page charges. We choose specific gauges where gauge potentials are well-defined on each coordinate patch and compute the Page charges. Using the integer $N$, introduced in the Page charge quantization with this gauge choice, we calculate the four-dimensional Newton's constant and the black hole entropy.
Unlike the Maxwell charges of the magnetic black holes, the Page charges do not reproduce the constraint satisfied by the parameters.

The paper is organized as follows. In section \ref{uplift-fomula}, we review the consistent truncations of the eleven-dimensional supergravity on a seven-dimensional Sasaki-Einstein manifold $M^{111}$ \cite{Cassani:2012pj}
and provide the uplifting formulae of four-dimensional AdS black holes \cite{Halmagyi:2013sla} to eleven-dimensions. In section \ref{explicit-sol}, we uplift the explicit near horizon solutions of black holes and compare them with previously known AdS$_2$ solutions in M-theory.
In section \ref{off-shell}, we present the off-shell identifications between the uplifting formulae and AdS$_2$ solutions. Then, we provide the off-shell expressions for black hole entropy.  
We discuss the flux quantization of Page charges in section \ref{page-charge}. We conclude in section \ref{discussion}.
In appendix \ref{ads2-review}, we review AdS$_2$ solutions in $D=11$ supergravity \cite{Gauntlett:2006ns, Donos:2008ug}.
In appendix \ref{Q111-HPZ}, we summarize the near horizon solutions of black hole asymptotic to AdS$_4 \times Q^{111}$ studied in \cite{Halmagyi:2013sla}.
In appendix \ref{page-detail}, we provide the detailed calculation of Page charges.
In appendix \ref{Newton's}, we discuss Newton's constant in four- and two-dimensions.
\section{Uplifting AdS black holes to eleven-dimensions}\label{uplift-fomula}
 In this paper, we revisit AdS black holes in four-dimensional gauged supergravity theory \cite{Halmagyi:2013sla}, which is obtained by consistent truncations of eleven-dimensional supergravity on seven-dimensional homogeneous manifolds with $SU(3)$-structures such as $Q^{111},\, M^{111},\, V_{5,2}$, etc. \cite{Cassani:2012pj}. 
 Here, let us focus on $Q^{111},\, M^{111}$. Then, the resulting four-dimensional theory is $\mathcal{N}=2$ gauged supergravity coupled to two ($Q^{111}$)/one ($M^{111}$) massless vector multiplets, a massive vector multiplet and one hypermultiplet, respectively. The massless vector field which is originated from the reduction of three-form flux on the non-trivial two-cycle of the internal manifold is called Betti vector field and corresponds to baryonic symmetry in dual field theory.
 
The AdS black hole solutions in these $\mathcal{N}=2$ gauged supergravity theories were studied in \cite{Halmagyi:2013sla}. They are static, supersymmetric and asymptotically AdS$_4$ black holes with electric and magnetic charges. Their near horizon geometry has a form of AdS$_2 \times \Sigma_g$ where $\Sigma_g$ is two-dimensional Riemann surface with genus $g \neq 1$.
The full black hole solutions interpolating AdS$_4$ and AdS$_2 \times \Sigma_g$ were studied only numerically. However, the near horizon solutions are known analytically and can be written in terms of the values of vector multiplet scalars at the horizon. In this note, we restrict ourselves to the near horizon solutions of AdS black holes.
 
In \cite{Halmagyi:2013sla}, it was already noted that the embedding of these AdS black holes in M-theory correspond to known AdS$_2$ solutions constructed by using wrapped branes \cite{Gauntlett:2006ns, Donos:2008ug}. For the case of magnetically charged black holes which asymptote to AdS$_4 \times M^{111}$, the metric uplift was explicitly studied in \cite{Kim:2019umc} and shown to be the same as that of AdS$_2$ solutions in M-theory with the electric four-form flux \cite{ Gauntlett:2006ns, Azzurli:2017kxo}. We extend this analysis to the fluxes. Then, we uplift the dyonic black hole solutions.

\subsection{Review of uplifting formulae}
We begin by reviewing the uplifting formulae for the metric and the four-form flux from the consistent truncation studied in \cite{Cassani:2012pj}.
The eleven-dimensional metric ansatz is
\be
ds_{11}^2= e^{2V} \mathcal{K}^{-1}ds_{4}^2+e^{-V}ds^2(B_6)+e^{2V}(\theta+A^0)^2.
\ee
The seven-dimensional internal space is written as a $U(1)$-fibration over a six-dimensional base $B_6$. This space is characterized by $SU(3)$-structure i.e., a real one-form $\eta=e^V \theta$, a real two-form $J$ and a complex three-form $\Omega$. Here $\mathcal{K}$ is related to the K\"ahler potential $K$ as $\mathcal{K}\equiv e^{-K}/8
$. From now on, we focus on $M^{111}$ case where $B_6$ is a product space of $\mathbb{CP}^2$ and $S^2$. Then the metric is written as
\be
ds_{11}^2= e^{2V} \mathcal{K}^{-1}ds_{4}^2+ \dfrac{1}{8} e^{2U_1} ds^2(\mathbb{CP}^2)+ \dfrac{1}{8}e^{2U_2} ds^2(S^2)+e^{2V}(\theta+A^0)^2.
\ee
Here, we normalized the scalar curvature of $S^2$ and $\mathbb{CP}^2$ to be $R(S^2)=2$ and $R(\mathbb{CP}^2)=4$.
The one-form $\theta$ has a form of $\theta=d\psi+\eta$ where $\psi$ has periodicity $\pi/2$ and
$d\eta=(J_{\mathbb{CP}_2}+ J_{S^2})/4$.
The 4D scalar fields $V,\, U_1,\, U_2$ are organized into $v_1,\, v_2,\, \phi$ as
\be
v_1 = e^{2U_1+V}, \quad v_2= e^{2U_2+V}, \quad \phi= -2U_1 -U_2.
\ee 
We also have $\mathcal{K}=v_1^2\,v_2$. Then, the eleven-dimensional metric is written as
\be\label{metric1}
ds_{11}^2=e^{\frac{4}{3}\phi} (v_1^2 v_2)^{-\frac{1}{3}}
\biggl(ds_4^2+v_1^2 v_2(\theta+A^0)^2+ \dfrac{e^{-2\phi} v_2}{8}ds^2(S^2)+\dfrac{e^{-2\phi} v_1}{8}ds^2(\mathbb{CP}^2) 
\biggr).
\ee
From the viewpoint of four-dimensional $\mathcal{N}=2$ gauged supergravity, $v_1, v_2$ are the imaginary part of the scalars in two vector multiplets and $\phi$ is the one of scalars in the hypermultiplet. $A_0$ is a graviphoton.

Now let us turn to the four-form flux in eleven-dimensional supergravity. The ansatz for three-form potential has a form of
\be\label{A3ansatz}
A_3 =C_3+B \wedge (\theta+A^0)-A^i \wedge w_i
+\xi^A \alpha_A-\tilde{\xi}_A \beta^A +b^i w_i \wedge(\theta+A^0).
\ee
Here $C_3$ is a three-form, $B$ is a two-form, $A_i$ are $n_V$ one-forms.
$w_i$ are $n_V$ two-forms and $\alpha_A ,\beta^A$ are  $2 n_H$ three-forms in $B_6$. 
For $M^{111}$ case, we have $i=1,2$ and $A=1$. 
The real scalars $b_i$ are the real part of the scalars in the vector multiplets.
The real scalars $\xi^A, \tilde{\xi}_A$ and scalar $a$ dual to three-form $B$
are the hypermultiplet scalars.
For the black hole solutions \cite{Halmagyi:2013sla} we are interested in, there is only one nontrivial hypermultiplet scalar $\phi$, whereas the other scalars $\xi=\tilde\xi=0$ and $a$ can be set to zero.
 Then the four-form flux reduces to
\begin{align}
F_4&=dC_3-dA^i \wedge w_i+b^i w_i \wedge(d\theta+dA^0),\nn\\
&=dC_3+(-F^i +b^i F^0)\wedge w_i+4(b^1+b^2)\tilde{w}^1+ 4 b^1 \tilde{w}^2,
\end{align}
where 
\begin{align}
dC_3&= \mathcal{K}^{-1} e^{4\phi} (b^I e_I+\frac{1}{2} \mathcal{K}_{ijk}
   m^i b^j b^k) *_4 1,\nn\\
   &= \mathcal{K}^{-1} e^{4\phi} (b^0 e_0+ 4b_1 b_2+ 2b_1^2) *_4 1,
\end{align}
with $I=0,1,2$.
Here the two-, four-forms are defined as
\begin{align}
w_1 &=\dfrac{1}{8} J_{\mathbb{CP}_2}, \quad w_2 =\dfrac{1}{8} J_{S^2},\nn\\
\tilde{w}_1 &= \frac{1}{2} w_1 \wedge w_2=\dfrac{1}{128} J_{\mathbb{CP}_2} \wedge J_{S^2},\nn\\
\tilde{w}_2 &=\frac{1}{2} w_1 \wedge w_1=\dfrac{1}{128} J_{\mathbb{CP}_2} \wedge J_{\mathbb{CP}_2},
\end{align}
and they satisfy the following relations
\be
d\theta=2(w_1+w_2),\quad dw_1=dw_2=0.
\ee 
We also used the electric and magnetic gauging parameters as
\be
e_I=(e_0,0,0)\quad \textrm{and} \quad m^I=(0,2,2).
\ee
 The coefficient $\mathcal{K}_{ijk}$ is defined as
\be
\mathcal{K}=\dfrac{1}{6}\mathcal{K}_{ijk} v^i v^j v^k.
\ee
Since the values of the scalar fields are constant on the horizon,
we are interested in a class of solutions satisfying
\be
db^i=0.
\ee
Then, the final expression of the four-form flux is given by\footnote{Throughout this paper, we have used $ b_0=1$. The Freund-Rubin parameter $e_0$ is related to the $AdS_4$ radius and the non-trivial scalar fields at the AdS${}_4$ vacuum \cite{Halmagyi:2013sla} as
\be
R_{AdS_4}=\dfrac{1}{2} \left( \dfrac{e_0}{6}\right)^{3/4}, \quad v_i= \sqrt{\dfrac{e_0}{6}},\quad e^{-2\phi}=\dfrac{e_0}{6}.
\ee
Here we set $e_0=6$ for simplicity.}
\begin{align}\label{flux1}
F_4&= \dfrac{e^{4\phi}}{v_1^2 v_2} \left(6+ 4b_1 b_2+ 2b_1^2\right) *_4 1
+ \dfrac{1}{8}(-F^1 +b_1 F^0)\wedge J_{\mathbb{CP}_2}
+ \dfrac{1}{8}(-F^2 +b_2 F^0)\wedge J_{S^2}\nn\\
&\phantom{=}+\dfrac{b_1+b_2}{32} J_{\mathbb{CP}_2} \wedge J_{S^2}+   \dfrac{b_1}{32} J_{\mathbb{CP}_2} \wedge J_{\mathbb{CP}_2}.
\end{align}
\subsection{Uplifting near horizon solutions of AdS black holes}
Now let us consider the uplifting of the near horizon solutions \cite{Halmagyi:2013sla} of AdS black holes.
The four-dimensional space-time metric has a form of
\be\label{metric-ansatz}
ds_4^2= R_1^2\, ds^2(\textrm{AdS}_2)+R_2^2\, ds^2(\Sigma_g).
\ee
Then, the eleven-dimensional metric \eqref{metric1} becomes
\begin{align}
\label{metric2}
ds_{11}^2=&e^{\frac{4}{3}\phi} (v_1^2 v_2)^{-\frac{1}{3}}R_1^2
\biggl(ds^2(\textrm{AdS}_2)+\dfrac{v_1^2 v_2}{R_1^2}(\theta+A^0)^2\nn\\
&+ \dfrac{e^{-2\phi} v_2}{8 R_1^2}ds^2(S^2)+\dfrac{e^{-2\phi} v_1}{8 R_1^2}ds^2(\mathbb{CP}^2) 
+ \dfrac{R_2^2}{R_1^2}ds^2(\Sigma_g)\biggr).
\end{align}
The solutions in \cite{Halmagyi:2013sla} i.e., $R_1^2,\, R_2^2,\, \phi$ are written in terms of $v_1$ and $v_2$.
We will substitute the explicit solutions into this uplifting formula \eqref{metric2} and write down the AdS$_2$ solutions in eleven-dimensional supergravity in the next section. 

For the uplifting of the four-flux, let us begin with the ansatz as
\be
F^I =-\tilde{q}^I_0\, R_1^2 \textrm{vol}(AdS_2) - p^I\, \kappa\, \textrm{vol}(\Sigma_g),
\ee
where $\kappa=1$ for $S^2$ and $\kappa=-1$ for $H_2$.
Then, the dual field strength becomes
\begin{align}
G_I &= \mathcal{R}_{IJ}F^J -\mathcal{I}_{IJ}* F^J,\nn\\
    &= \dfrac{R_1^2}{R_2^2}\Big(-R_2^2\, \mathcal{R}_{IJ}\,\tilde{q}^J_0 -\mathcal{I}_{IJ} p^J \kappa \Big)\textrm{vol}(AdS_2) -q_I\textrm{vol}(\Sigma_g).
\end{align}
Here $\mathcal{R}_{IJ},\mathcal{I}_{IJ}$ are the real, imaginary part of the period matrix $\mathcal{N}_{IJ}$
\be
\mathcal{N}_{IJ}=\bar{\mathcal{F}}_{IJ}+2\,i \dfrac{\textrm{Im}{\mathcal{F}}_{IK}\textrm{Im}{\mathcal{F}}_{JL}X^K X^L}{\textrm{Im}{\mathcal{F}}_{IJ}X^I X^J},
\ee
where $\mathcal{F}_{IJ}=\dfrac{\partial \mathcal{F}}{\partial X^I \, \partial X^J}$. $\mathcal{F}$ is the prepotential and $X^I$ are the homogeneous coordinates.
The electric and magnetic charges can be obtained by 
\begin{align}
p^I \kappa &=\dfrac{1}{4\pi} \int_{S^2} F^I,\nn\\
q_I &=\dfrac{1}{4\pi} \int_{S^2} G_I= -R_2^2\, \mathcal{I}_{IJ}\, \tilde{q}_0^J+ \mathcal{R}_{IJ}\,\kappa\,p^J.
\end{align}
In the consistent truncations investigated by Cassani, Koerber and Varela (which we abbreviate to CKV) in \cite{Cassani:2012pj}, they worked out in the frame where the electric and magnetic gauging parameters are
\be\label{CKV-gauging}
e_I=(e_0,0,0)\quad \textrm{and} \quad m^I=(0,2,2).
\ee
In this frame, the prepotential and the homogeneous coordinates are following
\be
\mathcal{F}=-\dfrac{X^1 X^2 X^3}{X^0}, \quad X^I=(1, z_1, z_2, z_3).
\ee
However, the four-dimensional black hole solutions studied by  Halmagyi, Petrini and Zaffaroni (which we abbreviate to HPZ) in \cite{Halmagyi:2013sla}, the authors worked out in the purely electric gauging as follows
\be
e_I=(e_0,2,2)
\footnote{In \cite{Halmagyi:2013sla}, they used $m^i=-2$ for dyonic gauging and $e_i=-2$ for electric gauging.
}
\quad \textrm{and} \quad m^I=(0,0,0).
\ee
The prepotential and the homogeneous coordinates are
\be
\mathcal{F}=2\sqrt{X^0 X^1 X^2 X^3}, \quad X^I=(1, z_2 z_3, z_3 z_1, z_1 z_2).
\ee
The purely electrically gauged frame (HPZ) can be obtained by a symplectic rotation of the dyonically gauged frame (CKV) as follows
\be
\begin{pmatrix} m^I  \\ e_I \end{pmatrix}_{HPZ}=S\begin{pmatrix} m^I  \\ e_I \end{pmatrix}_{CKV},
\ee
where
\be
S=\begin{pmatrix} A & B \\C & D \end{pmatrix}
 ,\quad A=D=\textrm{diag}(1,0,0,0),\quad B=-C=\textrm{diag}(0,-1,-1,-1).
\ee
Under a symplectic rotation, 
$\left(X^I, \, \mathcal{F}_I \right)$ and $\left( F^I, \, G_I \right)$ also transform.
More specifically, the field strengths in the two different frames are related as
\be
\begin{pmatrix} F^0 \\ F^i \\ G_0 \\G_i \end{pmatrix}_{CKV}
=\sqrt{2}\begin{pmatrix}  F^0 \\ G^i \\ G_0 \\-F_i  \end{pmatrix}_{HPZ}.\footnote{From now on, we include a numerical factor of $\sqrt{2}$. See the footnote 9 in \cite{Halmagyi:2013sla}.}
\ee
As a result, the four-form flux \eqref{flux1} becomes
\begin{align}\label{flux-uplift}
F_4
&= \dfrac{e^{4\phi}}{v_1^2 v_2}R_1^2\, R_2^2 \left(6+ 4b_1 b_2+ 2b_1^2\right) 
\, \textrm{vol}(AdS_2) \wedge \textrm{vol} (\Sigma_g)\nn\\
&\phantom{=}+ \dfrac{\sqrt{2}}{8}(-G^1_{HPZ} +b_1 F^0_{HPZ})\wedge J_{\mathbb{CP}_2}
+ \dfrac{\sqrt{2}}{8}(-G^2_{HPZ} +b_2 F^0_{HPZ})\wedge J_{S^2}\nn\\
&\phantom{=}+\dfrac{b_1+b_2}{32} J_{\mathbb{CP}_2} \wedge J_{S^2}+   \dfrac{b_1}{32} J_{\mathbb{CP}_2} \wedge J_{\mathbb{CP}_2}.
\end{align}
\section{Uplifted AdS$_2$ solutions in $D=11$ supergravity}\label{explicit-sol}
In this section, we consider the explicit near horizon solutions of black holes studied in \cite{Halmagyi:2013sla} and uplift them to the eleven-dimensions. Let us briefly summarize the near horizon solutions of dyonically charged black hole asymptotic to AdS$_4 \times M^{111}$ \cite{Halmagyi:2013sla}. There are two unknown function: the AdS radius $R_1$ and the size $R_2$ of two-dimensional Riemann surface $\Sigma_g$. We also have two complex scalars $b_i+i v_i$ in vector multiplets and one hypermultiplet scalar $\phi$. The black holes have electric charges $q_0,\, q_i$ and magnetic charges $p_0,\,p_i$.
The solutions are described in terms of two parameters $v_1,\, v_2$. We write down the solutions for readers' convenience.
\begin{align}\label{M111sol}
R_1^2 &=\dfrac{v_1 ^2 v_2}{16}, \quad
R_2^2 = \kappa R_1 ^2 \dfrac{2v_1^4+8v_1^3 v_2+(3e_0+8v_1^2)v_2^2}{v_2(3e_0 v_2-4v_1(v_1+2v_2)^2)},\\
e^{2\phi}&=\dfrac{4(v_1+2v_2)^2}{2v_1^4+8 v_1^3 v_2+(3e_0+8v_1^2)v_2^2},\quad
b_1= \epsilon_2 \sqrt{\dfrac{v_1(e_0-2v_1(v_1+2v_2))}{2(v_1+2v_2)}}, \quad
b_2 =-\dfrac{(v_1+v_2)b_1}{v_1}.\nn
\end{align}
We substitute the above solutions $e^{2\phi}, R_1, R_2 $ into the eleven-dimensional metric \eqref{metric2} and obtain 
\begin{align}\label{metric-sol} 
ds_{11}^2=&L^2\biggl(ds^2(\textrm{AdS}_2)+16(\theta+\sqrt 2 A^0)^2 
+(v_1^4+4 v_1^3 v_2+9 v_2^2+ 4v_1^2 v_2^2)
\biggl(\dfrac{1}{v_1^2(v_1+2v_2)^2}ds^2(S^2)\nn \\
&+\dfrac{1}{v_1 v_2(v_1+2v_2)^2}ds^2(\mathbb{CP}^2)+\dfrac{-\kappa}{v_2(2v_1^3-9v_2+8v_1^2 v_2+8 v_1 v_2^2)} ds^2(\Sigma_g)\biggr)\biggr),
\end{align}
where
\be
L^3=\dfrac{v_1^2 v_2 (v_1+2v_2)^2}{32(v_1^4+4 v_1^3 v_2+9 v_2^2+4 v_1^2 v_2^2)}.
\ee
For the four-form flux \eqref{flux-uplift}, the solution $b_1,\, b_2$ are also needed. Since the explicit expression of the flux is quite messy, we will not write down here.

The Bekenstein-Hawking entropy of the black hole can be easily calculated from the area law as
\be
S_{\textrm{BH}}= \dfrac{\textrm{Area}}{4 G_N^{(4)}}= \dfrac{\pi}{ G_N^{(4)}}|g-1| R_2^2
=\dfrac{1}{G_N^{(4)}} \pi  |g-1| \kappa\dfrac{v_1^2\Big(v_1^4+4v_1^3 v_2+(9+4v_1^2)v_2^2\Big)}{16\Big(9 v_2-2 v_1(v_1+2v_2)^2\Big)}.
\ee
The black hole entropy is expressed in terms of the vector multiplet scalars $v_1$ and $v_2$. Since the magnetic and electric charges of the black holes also can be written by using these scalar fields, one can write down the entropy in terms of black hole charges, in principle. However, the expressions are quite non-linear and not easily invertible apparently. One can also reproduce the entropy by calculating the two-dimensional Newton's constant following the method described in \cite{Couzens:2018wnk}.

In \cite{Azzurli:2017kxo}, the authors argued that the entropy can be written by using the three-sphere free energy as\footnote{See the equation (4.24) in \cite{Azzurli:2017kxo}.}
\begin{align}\label{ABCMZ-ent}
S_{\textrm{BH}}&= \dfrac{u}{8\pi} \textrm{vol}_{\Sigma_g}F_{S^3},
\end{align}
where $u $ corresponds to $R_2^2/R_1^2$ in our convention.
While the Bekenstein-Hawking entropy is given by
\be
S_{\textrm{BH}}=\dfrac{1}{4G_N^{(4)}}\textrm{vol}_{\Sigma_g}\, R_2^2,
\ee
the on-shell action of Euclidean AdS$_4$ whose boundary is $S^3$ is 
\be
F_{S^3}=\dfrac{\pi}{2} \dfrac{1/4}{G_N^{(4)}},
\ee
where the radius of vacuum AdS$_4$ is $1/2$.
Then, we obtain
\be\label{ent-F3}
S_{\textrm{BH}}= \dfrac{2}{\pi} \textrm{vol}_{\Sigma_g}\, R_2^2\, F_{S^3}=\dfrac{2}{\pi} \textrm{vol}_{\Sigma_g}\, u\, R_1^2\, F_{S^3}.
\ee
 For the black holes with a universal twist,\footnote{We can turn off the baryonic charges by setting $v_1=v_2=1$ and consider a universal twist.} the value of $R_1$ becomes $1/4$ and the equation \eqref{ABCMZ-ent} holds. But, for the general solutions, $R_1$ is a non-trivial function of $v_1$ and $v_2$ \eqref{M111sol}. Hence, our expression \eqref{ent-F3} does not match with \eqref{ABCMZ-ent}. 

In the subsequent sections, 
we consider the uplifting of the magnetic and dyonic black holes separately and compare them to the known AdS$_2$ solutions in the literatures.
The AdS$_2$ solutions in M-theory was classified by considering M2-branes wrapped on two-cycles in Calabi-Yau five-folds in \cite{Kim:2006qu}. The nine-dimensional internal space is a $U(1)$-fibration over the eight-dimensional K\"ahler manifold. We focus on a particular class of solutions where the eight-dimensional K\"ahler manifolds are products of two-dimensional K\"ahler-Einstein manifolds \cite{Gauntlett:2006ns, Donos:2008ug}. Recently, these solutions were revisited in \cite{Azzurli:2017kxo, Hong:2019wyi}. We will show that the uplifts of magnetic, dyonic black holes correspond to AdS$_2$ solutions in M-theory with electric and dyonic baryonic fluxes, respectively, studied in  \cite{Gauntlett:2006ns, Donos:2008ug, Azzurli:2017kxo, Hong:2019wyi}. 
\subsection{Uplifting magnetic black holes}\label{MBH}
First, let us consider a simple case, i.e. the magnetically charged black holes.
The electric charges of the black holes can be turned off by setting $v_2= (3-v_1^2)/(2v_1)$. In this case, the solutions can be written in terms of one parameter $v_1$ only. The metric uplift was discussed in the appendix of  \cite{Kim:2019umc}. From \eqref{metric-sol}, we have
\begin{align}\label{metric-mag}
ds_{11}^2=&\dfrac{1}{4}\left( \dfrac{v_1(3-v_1^2)}{2(9-2v_1^2+v_1^4)}\right)^{2/3}\biggl(ds^2(\textrm{AdS}_2)+16(\theta+\sqrt 2 A^0)^2 \\
&+(9-2v_1^2+v_1^4)
\left(\dfrac{1}{4v_1^2}ds^2(S^2)+\dfrac{1}{6-2v_1^2}ds^2(\mathbb{CP}^2)+\dfrac{1}{3+2v_1^2-v_1^4} ds^2(\Sigma_g)\right)\biggr).\nn
\end{align}
The uplift of flux \eqref{flux-uplift} is very simple for the magnetically charged black hole since the real part of the vector multiplet scalars $b_i$ vanish.
\be
F_4= \dfrac{e^{4\phi}}{v_1^2 v_2} 6\, R_1^2\, R_2^2\, \textrm{vol}(AdS_2) \wedge \textrm{vol} (\Sigma_g)
-\dfrac{\sqrt{2}}{8}\,G^1_{HPZ} \wedge J_{\mathbb{CP}_2}
-\dfrac{\sqrt{2}}{8}\,G^2_{HPZ} \wedge J_{S^2}\label{flux-MBH}.
\ee
We compute the dual field strengths as
\begin{align}\label{dualFS}
G^1_{HPZ}&=\frac{1}{v_2}*F^1=\frac{1}{v_2} \left(\dfrac{R_1^2}{R_2^2}\,p^1 \kappa\,\textrm{vol}(AdS_2)\right),\nn\\
G^2_{HPZ}&=\frac{v_2}{v_1^2}*F^2=\frac{v_2}{v_1^2} \left(\dfrac{R_1^2}{R_2^2}\,p^2 \kappa\,\textrm{vol}(AdS_2)\right).
\end{align}
Here we used
\be
 \mathcal{R}_{IJ}=0,\quad \mathcal{I}_{IJ}=\textrm{diag}\left(-v_1^2\,v_2,\,-\frac{1}{v_2},\,-\frac{v_2}{v_1^2}\right),
\ee
and
\be
q_I=0,\quad \tilde{q}_0^I=0,
\ee
for the magnetic black holes. Then the explicit form of the four-form flux becomes
\be\label{flux-mag}
F_4= L^3 \textrm{vol}(AdS_2) \wedge \left(
\dfrac{-12\kappa}{(3-v_1^2)(1+v_1^2)}J_{\Sigma_g}
+\dfrac{3+v_1^4}{2(3-v_1^2)} J_{\mathbb{CP}^2}
+\dfrac{(3-v_1^2)^2}{4v_1^2}J_{S^2}\right).
\ee
We note that the {\it magnetic} black hole solutions in four-dimensions correspond to the $AdS_2$ solutions with {\it electric} flux in eleven-dimensions.
\subsubsection{Flux quantization}\label{fluxquant-sol}
 Before we proceed further, 
let us consider the various cycles in eleven-dimensional background we are interested in. For the detailed discussions, see, for example, \cite{Gauntlett:2019roi, Klebanov:2010tj,Fabbri:1999hw}.
 
We begin with two- and five-cycles in $M^{111}$. A seven-dimensional Sasaki-Einstein manifold $M^{111}$ is a $U(1)$-fibration over $S^2 \times \mathbb{CP}^2$.
The first two-cycle $\mathcal{S}_1$ is the two-sphere $S^2$ and the second two-cycle $\mathcal{S}_2$ is obtained by fixing a point in $S^2 \subset\mathbb{CP}^2$. 
Since the K\"ahler form $J=J_{S^2}+J_{\mathbb{CP}^2}$ is trivial
\be
J_{S^2}+J_{\mathbb{CP}^2}=4 d\left(D\psi\right),
\ee
we have the exact two-form $J$. Then two 2-cycles $\mathcal{S}_1$ and $\mathcal{S}_2$ are not independent. Using the relation
\be
\int_{\mathcal{S}} J=0,
\ee
we can find an independent two-cycle $\mathcal{S}$, which can be used as a basis of the second homology $H_2(M^{111}; \mathbb{Z})$ such that
\be
\left(m\,n_a \int_{\mathcal{S}_1} -2\int_{\mathcal{S}_2} \right) J=0.
\ee
Here we follow the convention of \cite{Gauntlett:2019roi}. The volumes of $S^2$ and $\mathbb{CP}^2$ are given by
\be
\textrm{Vol}(S^2)=4\pi, \quad \textrm{Vol}(\mathbb{CP}^2)=\dfrac{1}{2}(2\pi)^2 M= 18 \pi^2.
\ee
We also have
\be
\int_{S^2} J_{s^2}=4\pi,\quad \int_{\Sigma_a} J_{\mathbb{CP}^2}=2\pi\, m\,n_a=6\pi.
\ee
Here $\mathbb{CP}^2$ can be replaced by other K\"ahler-Einstein four-manifold, for example, $\mathbb{CP}^1 \times \mathbb{CP}^1$. The integers $M,\, m,\, n_a$ is determined by the topology  of $KE_4^+$.
 The data for $KE_4^+=\mathbb{CP}^2,\, \mathbb{CP}^1 \times \mathbb{CP}^1$ are summarized in a table below.
 \begin{center}
\begin{tabular}{c|| c c c c}
&$m$ & $M$ & $n_a$ &\\
\hline
$\mathbb{CP}^2$ &3 &9 & 1 & $a=1$ \\
$\mathbb{CP}^1 \times \mathbb{CP}^1$ &2 &8 & 1 & $a=1,2$
\end{tabular}
\end{center}
   $\Sigma_a$ are two-cycles in K\"ahler-Einstein four-manifold. For $KE_4^+=\mathbb{CP}^2$ case, we have $\Sigma_1=\mathcal{S}_2$. We refer to \cite{Gauntlett:2019roi} for details.   
   
 We also have the two five-cycles $\mathcal{H}_1, \mathcal{H}_2$ obtained by fixing a point in $S^2$ and $S^2 \subset\mathbb{CP}^2$ in $M^{111}$, respectively.
These five-cycles are not independent and satisfy the following relation\footnote{This relation can be easily obtained by using the toric geometry description i.e., $\sum_{a=1}^d v_a^i[T_a]=0 \in  H_5(Y_7 ; \mathbb{Z})$ where $T_a$ are the toric five-cycles and $v_a^i$ is given by the toric data
\cite{Gauntlett:2019roi}.}
\be
2 \mathcal{H}_1+ 3\mathcal{H}_2 =0,
\ee
which corresponds to a Poincar\'e dual to the exact two-form $J$.

Now we consider the four- and seven-cycles in eleven-dimensional background $AdS_2 \times Y_9$ where $Y_9$ is obtained by fibering $M^{111}$ space over $\Sigma_g$. The independent four-cycle $\tilde{H}$ can be obtained by multiplying $\Sigma_g$ to the two-cycle $\mathcal{S}$ as
\be
\tilde{H}\equiv m\,n_a\,\Sigma_g \times \mathcal{S}_1-2\,\Sigma_g \times \mathcal{S}_2 .
\ee
The seven-cycles $Y_7,\, \mathcal{C}_1,\,\mathcal{C}_2 $ can be obtained by fixing a point on $AdS_2$ and $\Sigma_g,\, S^2,\, S^2 \subset\mathbb{CP}^2$, respectively.
Then, we integrate the Hodge dual of four-form flux \eqref{flux-mag} over these seven-cycles and consider the flux quantization as\footnote{Here we used 
$\textrm{Vol}(\Sigma_g)=4\pi (g-1)$.}
\begin{align}\label{fluxovercycle}
\dfrac{1}{(2\pi l_p)^6}\int_{Y_7} *_{11}F_4 &=\left(\dfrac{L}{l_p}\right)^6 \dfrac{M}{(2\pi)^2}\dfrac{3(9-2v_1^2+v_1^4)^2}{4 v_1^2 (3-v_1^2)^2} \equiv
N,\\
\dfrac{1}{(2\pi l_p)^6}\int_{\mathcal{C}_1} *_{11}F_4 &=\left(\dfrac{L}{l_p}\right)^6 \dfrac{M}{(2\pi)^2}(g-1)
\dfrac{\sqrt{2} p_2(9-2v_1^2+v_1^4)^2}{v_1^2 (3-v_1^2)^2}
=(g-1)\dfrac{4\sqrt{2} }{3 }p_2 N\equiv N_1,\nn\\
\dfrac{1}{(2\pi l_p)^6}\int_{\mathcal{C}_2} *_{11}F_4 &=\left(\dfrac{L}{l_p}\right)^6 \dfrac{m\,n_a}{(\pi)^2}(g-1)
\dfrac{\sqrt{2} p_1(9-2v_1^2+v_1^4)^2}{v_1^2 (3-v_1^2)^2}
=(g-1)\dfrac{4\,m\,n_a}{M}\dfrac{4\sqrt{2} }{3}p_1 N\equiv N_2.\nn
\end{align}
Here $N,\, N_1,\, N_2$ are all integers. $p_1$ and $p_2$ are the magnetic charges of the black hole.
We note that these seven-cycles are not independent and satisfy the following relation\footnote{This relation in $Y_9$ can be obtained by 
$
\sum_{a=1}^d v_a^i[\Sigma_a]=-n_i [Y_7] \in  H_7(Y_9 ; \mathbb{Z})
$
where $\Sigma_a$ are the seven-cycles. The twisting parameter $\vec{n}$ is given by $2(1-g)(1,0,0,1)$. See \cite{Gauntlett:2019roi} for more detailed explanation.
}
\be\label{constraint-cycle}
2 \mathcal{C}_1 +3 \mathcal{C}_2 +2(1-g) Y_7=0.
\ee
We integrate the flux $*_{11}F_4$ on this cycle using \eqref{fluxovercycle} and obtain 
\be\label{dirac-m111}
2p_1+p_2= \dfrac{3 \sqrt{2}}{8},
\ee
which is a consequence of the Dirac quantization condition studied in \cite{Halmagyi:2013sla}. 

Now let us more elaborate on the independent seven-cycle associated to the Betti vector field. In addition to the exact two-form $J=J_{S^2}+J_{\mathbb{CP}^2}$ in $M^{111}$, there is a harmonic two-form $\omega=J_{\mathbb{CP}^2}-2J_{S^2}$. We rewrite the relevant part of the three-form potential \eqref{A3ansatz} in terms of two-form $J$ and $w$ as
\begin{align}
A_3&=-\dfrac{\sqrt{2}}{8}\left( A^1 \wedge J_{\mathbb{CP}^2}+ A^2 \wedge J_{S^2} \right )+\cdots,\nn\\
&= -\dfrac{\sqrt{2}}{24}\Big\{ \left(2 A^1 +A^2 \right) \wedge  \left( J_{\mathbb{CP}^2}+ J_{S^2}  \right)+\left( A^1 -A^2 \right) \wedge  \left( J_{\mathbb{CP}^2}-2 J_{S^2}  \right)\Big\}+\cdots.
\end{align}
These particular linear combinations of vector fields  in the consistent truncation of \cite{Cassani:2012pj} can be explained as follows. The vector $2A^1+A^2$ mixes with the graviphoton $A^0$ to give a massless and a massive vector field \cite{Monten:2016tpu}. The vector $A^1-A^2$ is nothing but the Betti vector field we are interested in.\footnote{For $Q^{111}$ case, the detailed discussions are given in \cite{Donos:2012sy}.
}
We find that its field strength is proportional to the difference of the magnetic charges $p_1-p_2$ by using the equation \eqref{dualFS}.\footnote{For the squashed $M^{111}$, a harmonic two-form is $\omega=J_{\mathbb{CP}^2}/\ell_2^2-2J_{S^2}/\ell_1^2$. The story goes similarly.} Then, the charge can be calculated by using \eqref{fluxovercycle} as
\be
p_2-p_1 \propto \left(4\,m\, n_a\int_{\mathcal{C}_1}-M \int_{\mathcal{C}_2}\right)*F_4.
\ee
Hence, we have a non-trivial seven-cycle $H$ associated to the Betti vector field as
\be\label{cycleH}
H\equiv 4\, m\,n_a\,\mathcal{C}_1 -M\,\mathcal{C}_2.
\ee

Now we calculate the electric and magnetic fluxes as
\begin{align}\label{ef-uplift}
\mathcal{F} & \equiv \int_{H} *F_4=-L^6 2^4\sqrt{2} m n_a\dfrac{(p_1-p_2)(9-2v_1^2+v_1^4)^2}{v_1^2 (-3+v_1^2)^2} \textrm{Vol}(\Sigma_g)\textrm{Vol}(\mathbb{CP}^2)\Delta_\psi,\nn\\
\tilde{\mathcal{F}} & \equiv \int_{\tilde{H}} F_4=0,
\end{align}
where the seven- and four-cycle for $M^{111}$ are
\begin{align}
H&\equiv 12\,\mathcal{C}_1 -9\,\mathcal{C}_2,\\
\tilde{H}&\equiv 3\,\Sigma_g \times \mathcal{S}_1-2\,\Sigma_g \times \mathcal{S}_2,
\end{align}
and the period of $\psi$ is $\Delta_\psi
=\pi/2$.
\subsubsection{Comparison to known AdS$_2$ solutions}
Now let us come back to the uplifting of magnetic black holes. In this section, we will show that these uplifted solutions correspond to the $AdS_2$ solutions in eleven-dimensional supergravity without magnetic four-form flux \cite{Gauntlett:2006ns}. Once we identify $v_1$ in the metric \eqref{metric-mag} and the flux \eqref{flux-mag} as
\be\label{vandl}
v_1=\sqrt{\dfrac{3\ell_1}{2+\ell_1}},
\ee
we reproduce the metric and the flux in \cite{Gauntlett:2006ns}\footnote{Here we denote $L$ used in \cite{Gauntlett:2006ns} by $L_{GKW}$. It is related to the AdS$_2$ radius $L$ we have used in the previous section as $L^3=L_{GKW}^3 e^{3A}$. As a result, we have $L_{GKW}^3 =\sqrt{3\ell_1/(2+\ell_1)}/32$.}
\begin{align}\label{GMS}
ds_{11}^2&=L_{GKW}^2 e^{2A} \left(ds^2(\textrm{AdS}_2)+(dz+P)^2+e^{-3A}
\left(\dfrac{1}{\ell_1}ds^2(S^2)+ds^2(\mathbb{CP}^2)+\dfrac{2+\ell_1}{1+2\ell_1} ds^2(\Sigma_g)\right)\right),\nn\\
F_4 &=L_{GKW}^3 \dfrac{2+\ell_1}{3+2\ell_1+\ell_1^2}\textrm{vol}(AdS_2) \wedge \left(
\dfrac{-(2+\ell_1)^2}{1+2\ell_1}\kappa J_{\Sigma_g}
+\dfrac{1+\ell_1+\ell_1^2}{2+\ell_1} J_{\mathbb{CP}^2}
+\dfrac{3}{\ell_1(2+\ell_1)} J_{S^2}\right),
\end{align}
where 
\be
e^{-3A}=\dfrac{3+2\ell_1+\ell_1^2}{2+\ell_1}.
\ee
Here we have compared to the explicit solutions in \cite{Gauntlett:2019roi}, instead of expressions in \cite{Gauntlett:2006ns}.
Let us compare the fluxes through the various seven-cycles \eqref{fluxovercycle} and those in \cite{Gauntlett:2019roi}.
Then, one can easily read off the magnetic charges of black holes in terms of $\ell_1$ as
\be
p_1=\dfrac{3}{4\sqrt 2}\dfrac{1+\ell_1+\ell_1^2}{(2+\ell_1)(1+2\ell_1)}, \quad 
p_2=\dfrac{9}{4\sqrt 2}\dfrac{\ell_1}{(2+\ell_1)(1+2\ell_1)}.
\ee
Using the identification \eqref{vandl}, these magnetic charges reduce to the solution of Halmagyi, Petrini and Zaffaroni \cite{Halmagyi:2013sla} as
\be\label{charge-HPZ}
p_1= \dfrac{3+v_1^4}{8 \sqrt{2} (1+v_1^2)}, \quad
p_2= \dfrac{v_1^2 (3-v_1^2)}{4\sqrt{2} (1+v_1^2)}.
\ee

These AdS$_2$ solutions in eleven-dimensional supergravity were revisited rather recently \cite{Azzurli:2017kxo}. As discussed in \cite{Hong:2019wyi}, the solutions with the electric baryonic charges studied in the section 4 of \cite{Azzurli:2017kxo} correspond to those of \cite{Gauntlett:2006ns}. With the following identifications
\be
u=\dfrac{3+2\ell_1+\ell_1^2}{1+2\ell_1},\quad v=\dfrac{3+2\ell_1+\ell_1^2}{\ell_1(2+\ell_1)},
\ee
one can check the explicit form of the metric and the four-form flux\footnote{Here, we follow the notation used in \cite{Azzurli:2017kxo}. See the equation (4.25) in that paper} 
\begin{align}\label{ABCMZ}
ds_{11}^2&=L^2 \Bigl( ds^2(\textrm{AdS}_2)+ u\, ds^2(\Sigma_g)+ v\, ds^2(S^2)+ \dfrac{4\,q\,u\,v}{v-u+u\,v}ds^2(\mathbb{CP}^2)\nn\\
&\quad\phantom{L^2 \Big(}+ (d\psi+2qA_B-A+y d\beta)^2\Bigr),\nn\\
F_4 &=L^3 \textrm{vol}(AdS_2) \wedge \left(
(u+1)J_{\Sigma_g}
+\dfrac{u-v+u v}{v-u+u v} J_{\mathbb{CP}^2}
+(v-1) J_{S^2}\right),
\end{align}
reduce to \eqref{GMS}.
The electric and magnetic fluxes are\footnote{When $\mathbb{CP}^2$ is replaced by $\mathbb{CP}^1 \times \mathbb{CP}^1$, this electric flux reduce to result of \cite{Azzurli:2017kxo} with  $H_{\textrm{here}}=8H_{\textrm{ABCMZ}},\, \psi_{\textrm{here}}=\psi_{\textrm{ABCMZ}}/4$. See the equation (4.28) in  that paper.}
\begin{align}\label{ef-abcmz}
\mathcal{F} & \equiv \int_{H} *F_4=-\dfrac{uv (u(v-1)-v)(u(v-3)+v)}{(u(v-1)+v)^2}L^6  2^4 m n_a
\textrm{Vol}(\Sigma_g)\textrm{Vol}(\mathbb{CP}^2)\Delta_\psi ,\\
\tilde{\mathcal{F}} & \equiv \int_{\tilde{H}} F_4=0.
\end{align}
Let us identify the electric fluxes of the uplifted solution \eqref{ef-uplift} and the solution \eqref{ef-abcmz}. Then we can easily read off the magnetic charges of the Betti vector field as
\be
p_1-p_2= \dfrac{3\sqrt{2}(-1+v_1^2)^2}{16(1+v_1^2)}.
\ee
It is consistent with the previous result \eqref{charge-HPZ}. See also the equation (5.21) in \cite{Kim:2019umc}.
\subsection{Uplifting dyonic black holes}\label{dyon-BH}
In this section, we study the uplift of the dyonically charged black hole solutions and identify them to known AdS$_2$ solutions in M-theory with the internal four-form flux studied in \cite{Donos:2008ug}. As in the case of magnetic black holes, these AdS$_2$ solutions were also revisited in \cite{Azzurli:2017kxo, Hong:2019wyi}. Because the form of the uplifted solutions, particularly the form of the flux, is quite messy, let us consider simple cases to make a comparison more explicitly.
 Then, we reproduce the solutions studied in the section 4.3.2 in \cite{Azzurli:2017kxo}. We will also reproduce these solutions along the line of \cite{Donos:2008ug} at the end of the appendix \ref{w4form}.
Now, we focus on the cases where the electric baryonic charge vanishes in the eleven-dimensional solutions.
 We note that we can turn off the electric baryonic charge by adjusting the ratio between the size of $S^2,\, \Sigma_g$ and $\mathbb{CP}^2$.\footnote{ As we will explain later, this is clearly seen from the formulation of \cite{Gauntlett:2006ns, Donos:2008ug}. More precisely, we can turn off the electric charge by setting $\ell_1=\ell_2$ for $\kappa=1$ or $\ell_2=-\ell_4$ for $\kappa=-1$. Here $\ell_1,\, \ell_2,\,\ell_4$ are related to the inverse radius of $S^2,\,\mathbb{CP}^2$ and $\Sigma_g$, respectively. See the equation \eqref{puremagnetic}.}
 
First, let us consider a case where
the ratio of the size of $S^2$ and $\mathbb{CP}^2$ in \eqref{metric-sol} becomes one.
Then we have $v_1=v_2,\, \kappa=1$. The metric \eqref{metric-sol} reduce to
\begin{align}
ds_{11}^2=L^2
\biggl( &ds^2(\textrm{AdS}_2)+16(\theta+\sqrt 2 A^0)^2\nn\\
&+ \dfrac{1+v_1^2}{v_1^2}\,ds^2(S^2)+\dfrac{1+v_1^2}{v_1^2}\,ds^2(\mathbb{CP}^2) 
+ \dfrac{1+v_1^2}{1-2v_1^2}\,ds^2(\tilde{S}^2)\biggr).
\end{align}
Here we used $\tilde{S}^2$ instead of $\Sigma_g$ for $\kappa=1$.
If we change a variable as $v_1\equiv 1/\sqrt{1+ w^2}$, then the metric and flux \eqref{flux-uplift} have a form of
\begin{align}\label{dyonic-kp-metric}
ds_{11}^2&=L^2
\biggl(ds^2(\textrm{AdS}_2)+16(\theta+\sqrt 2 A^0)^2\nn\\
&\phantom{=L^2\biggl(}+ (w^2+2)\,ds^2(S^2)+(w^2+2)\,ds^2(\mathbb{CP}^2) 
+ \dfrac{w^2+2}{w^2-1}\,ds^2(\tilde{S}^2)\biggr),\nn\\
F_4&=L^3 \textrm{vol}(AdS_2) \wedge \left( \dfrac{3}{w^2-1} J_{\tilde{S}^2}+(1+w^2)J_{\mathbb{CP}^2}+(1+w^2)J_{S^2}\right)\nn\\
&+\epsilon_2 L^3\Big[ w(2+w^2)J_{\mathbb{CP}^2} \wedge \left( J_{\mathbb{CP}^2}-J_{S^2}\right)
-\dfrac{w(2+w^2)}{w^2-1} J_{\tilde{S}^2} \wedge \left(J_{\mathbb{CP}^2}-2 J_{S^2} \right)\Big],
\end{align}
where $\epsilon_2=\pm$.
They successfully reproduce the known results. See the equation (4.32) of \cite{Azzurli:2017kxo} and the equation (3.25b) in \cite{Hong:2019wyi}, respectively.
The electric and magnetic fluxes are computed as
\begin{align}
\mathcal{F} & \equiv \int_{H} *F_4=0,\\
\tilde{\mathcal{F}} & \equiv \int_{\tilde{H}} F_4=-3\times 16\pi^2 \dfrac{\sqrt{2}}{8} \Big((q_1-q_2)-(b_1-b_2)p_0 \kappa \Big) 
=m n_a\times48\pi^2 L^3 \dfrac{w(2+w^2)}{w^2-1}\epsilon_2.\nn
\end{align}
The entropy of this black hole is
\begin{align}
S_{\textrm{BH}}&=
\dfrac{\pi}{16G_N^{(4)}}  \dfrac{2+w^2}{(w^2-1)(1+w^2)^{3/2}},\nn\\
&=\dfrac{1}{2} \dfrac{2+w^2}{(w^2-1)(1+w^2)^{3/2}}F_{S^3}.
\end{align}

Now, let us consider the second case where
the ratio of the size of $\Sigma_g$ and $\mathbb{CP}^2$ in \eqref{metric-sol} becomes one.
Then we have $v_2=(9-4v_1^2+3\sqrt{9-8v_1^2})/(8v_1),\,\kappa=-1$.
Now let us identify $v_1 \equiv \sqrt{1+3w^2}/(1+w^2)$ and we reporduce the metric and the flux as
\begin{align}\label{dyonic-km-metric}
ds_{11}^2&=L^2
\biggl(ds^2(\textrm{AdS}_2)+16(\theta+\sqrt 2 A^0)^2\nn\\
&\phantom{=L^2
\biggl(}+ \dfrac{2+3w^2}{1+3w^2}ds^2(S^2)+(2+3w^2)ds^2(\mathbb{CP}^2) 
+ (2+3w^2)ds^2(\Sigma_g)\biggr),\nn\\
F_4&=L^3 \textrm{vol}(AdS_2) \wedge \left( 3(1+w^2) J_{\Sigma^2}+(1+3w^2)J_{\mathbb{CP}^2}+\dfrac{1}{1+3w^2}J_{S^2}\right)\\
&+\epsilon_2 L^3\Big[w(2+3w^2)J_{\mathbb{CP}^2} \wedge \left( J_{\mathbb{CP}^2}-\dfrac{1}{1+3w^2}J_{S^2}\right)
-w(2+3w^2) J_{\Sigma_g} \wedge \left(J_{\mathbb{CP}^2}-\dfrac{2}{1+3w^2} J_{S^2} \right)\Big].\nn
\end{align}
See the equation (4.29) of \cite{Azzurli:2017kxo} and the equation (3.25b) in \cite{Hong:2019wyi}, respectively.\footnote{There are some errors in \cite{Azzurli:2017kxo} for $\kappa=-1$ case, which was corrected in \cite{Hong:2019wyi}. We compare the uplifted solution and \cite{Hong:2019wyi}, and check that they agree.}
The fluxes become
\begin{align}
\mathcal{F} & \equiv \int_{H} *F_4=0,\\
\tilde{\mathcal{F}} & \equiv \int_{\tilde{H}} F_4
=-3\times16\pi^2 \dfrac{\sqrt{2}}{8} \Big((q_1-q_2)-(b_1-b_2)p_0 \kappa \Big) 
=m n_a\times 48\pi^2 L^3 \dfrac{w(1+w^2)(2+3w^2)}{1+3w^2}\epsilon_2.\nn
\end{align}
The entropy of this black holes is
\begin{align}
S_{\textrm{BH}}&=  
\dfrac{\pi}{16G_N^{(4)}} |g-1| (2+3w^2)\dfrac{\sqrt{1+3w^2}}{(1+w^2)^{3}},\nn\\
&=\dfrac{1}{2}|g-1| (2+3w^2)\dfrac{\sqrt{1+3w^2}}{(1+w^2)^{3}}F_{S^3},
\end{align}
where $0< w< 1/\sqrt{3}$.

So far, we have considered simple cases where the electric baryonic charge vanishes. In general, we note that the uplift of {\it dyonic} four-dimensional black holes corresponds to the AdS$_2$ solution with {\it dyonic} baryonic charge in eleven-dimensions.
\section{Reinterpretation of AdS$_4$ black holes : Off-shell solutions}\label{off-shell}
In this section, we compare the uplift formulae of dyonic AdS$_4$ black holes and the AdS$_2$ solutions in M-theory studied in \cite{Gauntlett:2006ns, Donos:2008ug}. We identify all the quantities needed for describing four-dimensional black hole solutions, i.e., the metric, scalar fields and the electric and magnetic charges, with the parameters appearing in eleven-dimensional solutions.

As we have seen in the previous section, the near horizon solutions of dyonic AdS black holes associated with $M^{111}$ are described by using two scalar fields $v_1$ and $v_2$. On the other hand, the AdS$_2$ solutions in M-theory are written in terms of $\ell_i, m_{ij}$
as we will review in detail in the appendix \ref{ads2-review}. Once we impose the self-dual conditions \eqref{pri-con-1}, we have five parameters $\ell_1, \ell_2, \ell_4$ and $m_{12}, m_{14}$. Here we set $\ell_2=\ell_3, m_{12}=m_{13}$ for $M^{111}$. These parameters should satisfy one constraint \eqref{con-M111-2} as
\be
2\ell_1 \ell_2+\ell_1 \ell_4+\ell_2^2+2\ell_2 \ell_4
=2(2 m_{12}^2+m_{14}^2),
\ee
which is imposed by the equation of the motion, and the remaining primitivity condition \eqref{pri-con-2}
\be
2m_{12}+m_{14}=0.
\ee
Furthermore, there is a rescaling symmetry. Then, the number of the independent parameters are reduced from five to two, which is the same as the dimension of the solution space $(v_1, v_2)$.

We note that the AdS$_2$ solutions in \cite{Gauntlett:2006ns, Donos:2008ug} can be treated {\it off-shell} in a sense that one can write down the metric and flux without imposing the constraint 
\eqref{con-M111-2}, which is required by the equation of the motion. These off-shell description have two advantages. As we mentioned, the black hole solutions in \cite{Halmagyi:2013sla} are described by $v_1, v_2$ for $M^{111}$ case.  In this off-shell description, we have more parameters, i.e., $\ell_1, \ell_2, \ell_4, m_{12}, m_{14}$ before we impose the constraints and consider rescaling symmetry. It enables us to reinterpret the AdS black hole solutions in a more symmetric way. In addition to this, we obtain the expression for the entropy before imposing the equation of the motion.
It certainly follows the spirit of 
the extremization principles proposed in \cite{Couzens:2018wnk}, where one imposes the conditions for supersymmetry and relaxes the equation of the motion for the fluxes.

\subsection{Black holes asymptotic to AdS$_4 \times M^{111}$ }
In this section, we study the off-shell near horizon geometries of the black holes which are
asymptotically AdS$_4 \times M^{111}$. Based on this, we will generalize to the black holes asymptotic to AdS$_4 \times Q^{111}$ in the next section. 
\subsubsection{Magnetic black holes}\label{off-shell-mag}
Now let us compare the uplift formulae of the four-dimensional magnetically charged black holes, i.e., the metric \eqref{metric2} and the flux \eqref{flux-MBH} to AdS$_2$ solutions with electric flux in eleven-dimensional supergravity \eqref{metric-GKW-gen} and \eqref{flux-GKW-gen}. That allows us to identify the four-dimensional black hole solutions in terms of $\ell_i$s as
\begin{align}
R_1^2&=\dfrac{3\sqrt 3 \sqrt{\ell_1}\ell_2}{16(\ell_1+2\ell_2)^{3/2}},\quad
R_2^2=R_1^2 \dfrac{\ell_1+2\ell_2+\ell_4}{\ell_4 \kappa}
,\quad b_1=b_2=0,\nn\\
v_1&= \left(\dfrac{3\ell_1}{\ell_1+2\ell_2}\right)^{1/2},\quad
v_2= \left(\dfrac{3\ell_2^2}{\ell_1(\ell_1+2\ell_2)}\right)^{1/2},\quad
e^\phi= \left(\dfrac{2(\ell_1+2\ell_2)}{3(\ell_1+2\ell_2+\ell_4)}\right)^{1/2}.
\end{align}
The magnetic charges of the black holes are
\be\label{mag-ch-ell}
p^1 =-\dfrac{3\ell_2(\ell_1+\ell_2+\ell_4)}{4\sqrt 2 (\ell_1+2\ell_2)\ell_4}, \quad\quad
p^2 =-\dfrac{3\ell_1(2\ell_2+\ell_4)}{4\sqrt 2 (\ell_1+2\ell_2)\ell_4 }.
\ee
Here we have\footnote{The $L$ used in \cite{Gauntlett:2006ns} is denoted as $L_{GKW}$. It is related to $L$ used in \eqref{L-mag-m111} by $L_{GKW}^3=L^3 (\ell_1+2\ell_2+\ell_4)$.}
\be\label{L-mag-m111}
L^3=e^{2\phi} (v_1^2 v_2)^{-\frac{1}{2}} R_1^3=\dfrac{\sqrt{3\ell_1}\ell_2}{32(\ell_1+2\ell_2+\ell_4)\sqrt{\ell_1+2\ell_2}}.
\ee
This off-shell solution is eventually described by a single independent parameter among $\ell_1, \ell_2, \ell_4$.
When we set $\ell_2=1$ by rescaling and solve the constraint \eqref{constraint1}, we have 
\be
\ell_4=-\dfrac{1+2\ell_1}{2+\ell_1}.
\ee
Then the magnetic black hole solutions studied in the section \ref{MBH} can be reproduced with the identification as
\be
\ell_1\equiv-\dfrac{2v_1^2}{-3+v_1^2}.
\ee

The entropy of the magnetic black holes is given by
\be\label{SBH-m}
S_{\textrm{BH}}
=\dfrac{8}{3} \pi N^{3/2} |g-1| \dfrac{ \sqrt{\ell_1}\ell_2}{(\ell_1+2\ell_2)^{3/2}}
\dfrac{\ell_1+2\ell_2+\ell_4}{\ell_4 \kappa}.
\ee
When we consider rescaling and impose the constraint as above, we can reproduce the previously known results \cite{Gauntlett:2019roi, Kim:2019umc}.

Here, we provide a quite simple observation.
 In the study of the AdS black holes in $\mathcal{N}=2$ gauged supergravity \cite{Halmagyi:2013sla}, it is known that the Dirac quantization conditions give the constraints on the magnetic charges of the black hole as\footnote{See the equation \eqref{dirac-q} in the appendix for $Q^{111}$ case.}
 \be 
p_0=-\dfrac{1}{4\sqrt{2}}, \quad 2p_1+p_2=\dfrac{\sqrt 2 e_0}{16}.
\ee
Once we substitute the relation between the magnetic charges and the parameters $\ell_i$ \eqref{mag-ch-ell} into the second constraint above, we obtain
\be
2 \ell_1 \ell_2 +\ell_1 \ell_4+\ell_2^2 +2\ell_2 \ell_4=0,
\ee
which is nothing but the constraint \eqref{constraint1} required by the equation of motion. 
\subsubsection{Dyonic black holes}\label{dyon-M111}
As we have explained in the previous sections, the eleven-dimensional uplifts of the dyonic black holes correspond to AdS$_2$ solutions with non-vanishing internal four-form flux reviewed in the appendix \ref{w4form}. In addition to $\ell_i$, new parameters $m_{ij}$ are introduced due to the presence of the non-trivial internal four-form flux.
Here we only impose the self-duality conditions \eqref{pri-con-1} and identify $\ell_2=\ell_3,\,m_{12}=m_{13}$ for KE$_4$ case.
Then the black hole solutions can be expressed in terms of the five parameters $\ell_1, \ell_2, \ell_4, m_{12}$ and $m_{14}$.
\begin{align}\label{dictionary-M111-dyon}
R_1^2&=\dfrac{3\sqrt 3 \sqrt{\ell_1}\ell_2^4}{16(\ell_2^2(\ell_1+2\ell_2)
-4\ell_2 m_{12} m_{14}+\ell_1 m_{14}^2)^{3/2}},\quad
R_2^2=R_1^2 \dfrac{\ell_1+2\ell_2+\ell_4}{\ell_4 \kappa},\nn\\
e^\phi&= \left(\dfrac{2(\ell_2^2(\ell_1+2\ell_2)-4\ell_2m_{12}m_{14}+\ell_1 m_{14}^2)}{3\ell_2^2(\ell_1+2\ell_2+\ell_4)}\right)^{1/2},\nn\\
v_1&= \left(\dfrac{3\ell_1 \ell_2^2}{\ell_2^2(\ell_1+2\ell_2)
-4\ell_2m_{12}m_{14}+\ell_1 m_{14}^2}\right)^{1/2},\nn\\
v_2&= \left(\dfrac{3\ell_2^4}{\ell_1(\ell_2^2(\ell_1+2\ell_2)-4\ell_2m_{12}m_{14}+\ell_1 m_{14}^2)}\right)^{1/2},\nn\\
b_1&= 32 L^3\dfrac{\ell_1+2\ell_2+\ell_4}{\ell_2^2}m_{14},\nn\\
b_2&= -32 L^3\dfrac{\ell_1+2\ell_2+\ell_4}{\ell_1 \ell_2^2}(-2\ell_2 m_{12}+\ell_1 m_{14}),
\end{align}
where\footnote{The $L$ used in \cite{Donos:2008ug} is denoted as $L_{DGK}$. It is related to $L$ used in \eqref{L-dyon-m111} by $L_{DGK}^3=L^3 (\ell_1+2\ell_2+\ell_4)$.}
\be\label{L-dyon-m111}
L^3=e^{2\phi} (v_1^2 v_2)^{-\frac{1}{2}} R_1^3
=\dfrac{\sqrt{3\ell_1}\ell_2^2}{32(\ell_1+2\ell_2+\ell_4)\sqrt{\ell_2^2(\ell_1+2\ell_2)-4\ell_2m_{12}m_{14}+\ell_1 m_{14}^2}}.
\ee
The electric and magnetic charges of the black holes are written as\footnote{Here we used $\tilde{q}_0^0=0$.}
\begin{align}\label{charge-dyon-M111}
q^0&= -2 b_1 b_2 q^1-b_1^2 q^2+ \left(b_1^2 b_2 p^0 +2 b_1 p^1+b_2 p^2\right) \kappa,\nn\\
q^1&= 4\sqrt{2} L^3\dfrac{\ell_1+2\ell_2+\ell_4}{\ell_2^2 \ell_4}(2 \ell_2 m_{12}-\ell_4 m_{14} )\kappa,\nn\\
q^2&=4 \sqrt{2} L^3\dfrac{\ell_1+2\ell_2+\ell_4}{\ell_1 \ell_2^2 \ell_4}(2 \ell_2^2 m_{14}-2\ell_2 \ell_4 m_{12} +\ell_1 \ell_4 m_{14}) \kappa,\nn\\
p^0&=-\dfrac{1}{4\sqrt{2}},\nn\\
p^1  &= -\dfrac{3 \left( \ell_2^3 (\ell_1+\ell_2+\ell_4)-2 \ell_2(\ell_2 m_{14}^2+2\ell_2 m_{12}^2-\ell_1 m_{14}m_{12})+\ell_4 m_{14}(2\ell_2 m_{12}-\ell_1 m_{14}) \right)}{4\sqrt{2}\ell_4 \left(\ell_2^2 (\ell_1+2\ell_2)-4 \ell_2 m_{12}m_{14}+\ell_1 m_{14}^2 \right)},\nn\\
p^2 &= -\dfrac{3\ell_1 \left(\ell_2^2 (2\ell_2+\ell_4)-4 \ell_2 m_{14}m_{12}+\ell_4 m_{14}^2  \right)}{4\sqrt{2}\ell_4 \left(\ell_2^2 (\ell_1+2\ell_2)-4 \ell_2 m_{12}m_{14}+\ell_1 m_{14}^2 \right)}.
\end{align}
The solutions are eventually described by two independent parameters among $\ell_1, \ell_2, \ell_4$ and $m_{12}, m_{14}$ when we consider rescaling in $\ell_i$ and imposing the constraint \eqref{constraint2}, the remaining primitive condition \eqref{pri-con-2}. It is consistent to the result of \cite{Halmagyi:2013sla}, i.e. the general solution space of $M^{111}$ black hole is two-dimensional.

We calculate the entropy of dyonic black holes without imposing the equation of motion as
\begin{align}\label{SBH-d}
S_{\textrm{BH}}
&=\dfrac{\pi}{ G_N^{(4)}} |g-1|
\dfrac{3\sqrt 3 \sqrt{\ell_1}\ell_2^4}{16(\ell_2^2(\ell_1+2\ell_2)
-4\ell_2 m_{12} m_{14}+\ell_1 m_{14}^2)^{3/2}}
\dfrac{\ell_1+2\ell_2+\ell_4}{\ell_4 \kappa}.
\end{align}

Similarly for the case of magnetic black holes, the relations on the magnetic charges $2p_1+p_2=(\sqrt 2 e_0)/16$, which is a consequence of Dirac quantization, are translated into the equation of motion
\be
2 \ell_1 \ell_2 +\ell_1 \ell_4+\ell_2^2 +2\ell_2 \ell_4-2(2m_{12}^2+m_{14}^2)=0,
\ee
by using the relation \eqref{charge-dyon-M111}.

In \cite{Halmagyi:2013sla}, the authors presented the solutions associated to $M^{111}$ in terms of two scalars $(v_1, v_2)$. Alternatively, one can also consider an off-shell solutions in terms of four scalars $(v_i, b_i)$, which satisfy two constraints \eqref{BH-con1}, \eqref{BH-con2} as the authors did in $Q^{111}$ model. Using the dictionary \eqref{dictionary-M111-dyon}, we note that the first constraint \eqref{BH-con1} corresponds to the primitive condition
\be
2m_{12}+m_{14}=0,
\ee
while the second one \eqref{BH-con2} is satisfied automatically.
\subsection{Black holes asymptotic to AdS$_4 \times Q^{111}$}
So far, we have mainly focused on the asymptotically AdS$_4 \times M^{111}$ black holes. However, based on the results of the previous section, we can generalize the off-shell black hole solutions  to the $Q^{111}$ case by manifesting the symmetries between $\ell_1, \ell_2$ and $\ell_3$.
\subsubsection{Magnetic black holes}
For the magnetically charged black holes, we identify as follows
\begin{align}
R_1^2&=\dfrac{3\sqrt 3 \sqrt{\ell_1 \ell_2 \ell_3}}{16(\ell_1+\ell_2+\ell_3)^{3/2}},\quad
R_2^2=R_1^2 \dfrac{\ell_1+\ell_2+\ell_3+\ell_4}{\ell_4 \kappa}, \quad
e^\phi= \left(\dfrac{2(\ell_1+\ell_2+\ell_3)}{3(\ell_1+\ell_2+\ell_3+\ell_4)}\right)^{1/2},\nn\\
v_1&= \left(\dfrac{3\ell_2 \ell_3}{\ell_1(\ell_1+\ell_2+\ell_3)}\right)^{1/2},\quad
v_2= \left(\dfrac{3\ell_1 \ell_3}{\ell_2(\ell_1+\ell_2+\ell_3)}\right)^{1/2},\quad
v_3= \left(\dfrac{3\ell_1 \ell_2}{\ell_3(\ell_1+\ell_2+\ell_3)}\right)^{1/2},\quad\nn\\
p^1 &=-\dfrac{3\ell_1(\ell_2+\ell_3+\ell_4)}{4\sqrt 2 (\ell_1+\ell_2+\ell_3)\ell_4 }, \quad
p^2 =-\dfrac{3\ell_2(\ell_1+\ell_3+\ell_4)}{4\sqrt 2 (\ell_1+\ell_2+\ell_3)\ell_4}, \quad
p^3 =-\dfrac{3\ell_3(\ell_1+\ell_2+\ell_4)}{4\sqrt 2 (\ell_1+\ell_2+\ell_3)\ell_4}.
\end{align}
where 
\be
L^3=e^{2\phi} (v_1 v_2 v_3)^{-\frac{1}{2}} R_1^3=\dfrac{\sqrt{3\ell_1 \ell_2 \ell_3}}{32(\ell_1+\ell_2+\ell_3+\ell_4)\sqrt{\ell_1+\ell_2+\ell_3}}.
\ee
When we set $\ell_2=\ell_3$, we have $v_2=v_3$, we can reproduce the $M^{111}$ solution.\footnote{Note that the conventions used in the literatures \cite{Halmagyi:2013sla, Gauntlett:2019roi} are different. To reduce $Q^{111}$ solution to $M^{111}$, the authors in \cite{Gauntlett:2019roi} identified $\ell_2=\ell_3$. On the other hand, the authors in \cite{Halmagyi:2013sla} used the identification $v_1=v_3$. Hence, we should interchange the order of the vector multiplets when we reproduce $M^{111}$ solution using the result of this section. For example, we have $(v_1, v_2)_{\textrm{here}}=(v_2, v_1)_{\textrm{there}}$.}
\subsubsection{Dyonic black holes}
For the dyonically charged black holes,  we have only imposed the self-dual condition on $m_{ij}$ \eqref{pri-con-1} as $M^{111}$ case. Then we have
\begin{align}\label{myq111}
R_1^2&=\dfrac{3\sqrt 3 \ell_1^2\ell_2^2\ell_3^2}{16(\ell_1\ell_2\ell_3(\ell_1+\ell_2+\ell_3)
+\mathcal{M})^{3/2}},\quad
R_2^2=R_1^2 \dfrac{\ell_1+\ell_2+\ell_3+\ell_4}{\ell_4 \kappa},\nn\\
e^\phi&= \left(\dfrac{2(\ell_1\ell_2 \ell_3(\ell_1+\ell_2+\ell_3)
+\mathcal{M})}
{3\ell_1 \ell_2 \ell_3(\ell_1+\ell_2+\ell_3+\ell_4)}\right)^{1/2},\nn\\
b_i&= 32\, L^3\, \dfrac{\ell_1+\ell_2+\ell_3+\ell_4}{\ell_1 \ell_2 \ell_3} (\ell_1 m_{23}+\ell_2 m_{13}+\ell_3 m_{12}-2\ell_i m_{jk}),\nn\\
v_i&= \left(\dfrac{3\ell_j^2\ell_k^2}{\ell_1\ell_2 \ell_3(\ell_1+\ell_2+\ell_3)
+\mathcal{M}}\right)^{1/2},
\end{align}
where
\be
L^3=e^{2\phi} (v_1 v_2 v_3)^{-\frac{1}{2}} R_1^3
=\dfrac{\sqrt{3}\ell_1\ell_2 \ell_3}{32(\ell_1+\ell_2+\ell_3+\ell_4)\sqrt{\ell_1\ell_2\ell_3(\ell_1+\ell_2+\ell_3)
+\mathcal{M}}}.
\ee
Here we defined $\mathcal{M}$ as
\be\label{M-Q111}
\mathcal{M} \equiv
-2(\ell_1 m_{23}\, \ell_2 m_{31}+\ell_2 m_{31}\,\ell_{3} m_{12}+\ell_3 m_{12}\, \ell_1 m_{23})
+(\ell_1^2 m_{14}^2+\ell_2^2 m_{24}^2+\ell_3^2 m_{34}^2).
\ee
The electric and magnetic charges are
\begin{align}
q^0&=-(b_2 b_3 q_1+b_3 b_1 q_2 +b_1 b_2 q_3)+ (b_1 b_2 b_3 p_0+b_1 p_1 +b_2 p_2 +b_3 p_3) \kappa,\\
q^i&=4\sqrt{2}\,L^3\,\dfrac{\ell_1+\ell_2+\ell_3+\ell_4}{\ell_1 \ell_2 \ell_3 \ell_4} \Big( 2\ell_j \ell_k m_{jk}-\ell_4 (\ell_1 m_{23}+\ell_2 m_{13}+\ell_3 m_{12}-2\ell_i m_{jk})\Big)\kappa,\nn\\
p^0&=-\dfrac{1}{4\sqrt{2}},\nn\\
p^i&=-\frac{8}{\sqrt{2}}\,L^3\,\left(\dfrac{\ell_1+\ell_2+\ell_3+\ell_4}{\ell_4}\right) 
\left(\dfrac{\ell_1+\ell_2+\ell_3+\ell_4-\ell_i}{\ell_i}\right) \dfrac{v_j v_k}{v_i}-b_j b_k p_0+\kappa (b_j q^k+b_k q^j).\nn
\end{align}
Here we use $(i,\, j,\,k)=(1,\,2,\,3)$ in cyclic order. Note that there is no summation in the repeated indices.
Upon rescaling of $\ell_i$ and imposing the constraint \eqref{constraint2}, the remaining primitivity condition \eqref{pri-con-2} reduces the number of the independent parameters from 7 to 4. It is consistent to the black hole solution analysis in \cite{Halmagyi:2013sla} where it was shown that $Q^{111}$ black hole solutions are parameterized by $(b_i, v_i)$ where $i=1,2,3$ with two non-trivial constraints \eqref{BH-con1}, \eqref{BH-con2} between them. 
When we set $\ell_2=\ell_3$ and $m_{12}=m_{13}$, we have $v_2=v_3,\, b_2=b_3$ and reproduce the $M^{111}$ solution.

One can check that the solutions we provide in this section reduce to the solutions for $Q^{111}$ of Halmagyi, Petrini and Zaffaroni.  First, we substitute $v_i$ and $b_i$ \eqref{myq111} into the black hole solutions and charges summarized in the appendix \ref{Q111-HPZ}. Then their solutions can be also written in terms of $\ell_1,\,\ell_2,\,\ell_3$ and $m_{12},\,m_{23},\,m_{31}$. We have checked that they exactly agree with the solutions and charges in this section after we impose the constraint \eqref{constraint3} and eliminate $\ell_4$. 

The entropy of dyonic AdS$_4$ black holes for $Q^{111}$ is  
\begin{align}\label{SBH-d-Q111}
S_{\textrm{BH}}
=\dfrac{\pi}{ G_N^{(4)}} |g-1| 
\dfrac{3\sqrt{3}(\ell_1\ell_2\ell_3)^2}{16\left(\ell_1\ell_2\ell_3(\ell_1+\ell_2+\ell_3)
+\mathcal{M}\right)^{3/2}}\,\dfrac{\ell_1+\ell_2+\ell_3+\ell_4}{\ell_4 \kappa}.
\end{align}

\section{Page charge quantization and black hole entropy}\label{page-charge}
In this section we consider the flux quantization in AdS$_2$ solutions of eleven-dimensional supergravity studied in \cite{Gauntlett:2006ns, Donos:2008ug}, which we review in the appendix \ref{ads2-review}. There are two classes of AdS$_2$ solutions depending on whether the internal four-form fluxes vanish or not. 

First let us consider the case where the internal four-form fluxes vanish \eqref{flux-GKW-gen}. Then we integrate $*_{11}F_4$ over the various seven-cycles $Y_7, \mathcal{C}_1, \mathcal{C}_2$, which we studied in section \ref{fluxquant-sol}, and impose the flux quantization conditions as
\begin{align}\label{flux-cycle}
\dfrac{1}{(2\pi l_p)^6} \int_{Y_7} *_{11}F_4&=
\left(\dfrac{L_{GKW}}{l_p} \right)^6 \dfrac{\ell_1+2\ell_2}{\ell_1 \ell_2^2}\dfrac{9}{4 \pi^2}\equiv N,\nn\\
\dfrac{1}{(2\pi l_p)^6} \int_{\mathcal{C}_1} *_{11}F_4 &=
(g-1)\left(\dfrac{L_{GKW}}{l_p} \right)^6 \dfrac{2\ell_2+\ell_4}{\ell_2^2 \ell_4 \kappa}\dfrac{9}{4 \pi^2}\equiv N_1,\nn\\
\dfrac{1}{(2\pi l_p)^6} \int_{\mathcal{C}_2} *_{11}F_4 &=
(g-1)\left(\dfrac{L_{GKW}}{l_p} \right)^6 \dfrac{\ell_1+\ell_2+\ell_4}{\ell_1 \ell_2 \ell_4 \kappa}\dfrac{3}{ \pi^2}\equiv N_2.
\end{align}
Here, $N,\, N_1$ and $N_2$ are all integers.
As we mentioned, the seven-cycles $Y_7, \mathcal{C}_1, \mathcal{C}_2$ are not independent and satisfy the relation \eqref{constraint-cycle}
\be
2 \mathcal{C}_1 +3 \mathcal{C}_2 +2(1-g) Y_7=0.
\ee
Now we integrate $*_{11}F_4$ along this trivial-cycle and obtain
\be\label{flux-con-a}
2 N_1 +3 N_2 +\kappa 2(g-1) N=0,
\ee
which reduce to 
\be\label{eom-con-a}
2 \ell_1 \ell_2 +\ell_1 \ell_4+\ell_2^2 +2\ell_2 \ell_4=0,
\ee
by using the flux quantization conditions \eqref{flux-cycle}.
It is just the constraint equations on $\ell_i$ imposed by the equation of motion \eqref{constraint1}. We recall that this constraint also appears as the consequence of Dirac quantization conditions of four-dimensional black hole we explained at the end of the section \ref{off-shell-mag}.
Furthermore, 
when we identify $\mathbf{n_4}, \mathbf{n_5}$ as
\begin{align}\label{FQ}
\mathbf{n_5} \equiv -\dfrac{N_1}{N} =-(g-1) \dfrac{\ell_1(2\ell_2+\ell_4)}{(\ell_1+2\ell_2)\ell_4 \kappa},
\quad 
\mathbf{n_4} \equiv -\dfrac{N_2}{N} =-(g-1) \dfrac{4}{3}\dfrac{\ell_2(\ell_1+\ell_2+\ell_4)}{(\ell_1+2\ell_2)\ell_4 \kappa},
\end{align}
then, \eqref{flux-con-a} reduces to $3\mathbf{n}_4+2\mathbf{n}_5=\kappa2\,(g-1)$, which is a twisting condition. See, for example, \cite{Hosseini:2019use}.

These magnetic fluxes, together with the components of the Reeb vector field, appear in the expression of the entropy functional studied in \cite{Hosseini:2019use, Gauntlett:2019roi,Hosseini:2019ddy,  Kim:2019umc}. After extremizing with respect to the Reeb vector, the entropy functional for $M^{111}$ depends on the magnetic fluxes only and has a form of \footnote{See the equation (5.23) in \cite{Kim:2019umc}.}
\be\label{EF}
S=-\dfrac{\sqrt{2} \pi }{9} N^{3/2} \sqrt{\dfrac{9\mathbf{n}_4-\sqrt{81 \mathbf{n}_4^2-72 \mathbf{n}_4\mathbf{n}_5-48\mathbf{n}_5^2}}{3\left(3 \mathbf{n}_4+2 \mathbf{n}_5 \right)}} 
\left( 18\mathbf{n}_4+\sqrt{81 \mathbf{n}_4^2-72 \mathbf{n}_4\mathbf{n}_5-48\mathbf{n}_5^2}\right),
\ee
with $3\mathbf{n}_4+2\mathbf{n}_5=2(1-g)$.
We substitute the fluxes \eqref{FQ} into the entropy functional \eqref{EF} and obtain it in terms of $\ell_i$s. For the magnetic black holes, this expression agrees to the black hole entropy \eqref{SBH-m} when we set $\ell_2=1$ by rescaling and impose the constraint $\ell_4=-(1+2\ell_1)/(2+\ell_1)$ on both sides.\footnote{When $\ell_1>1$, two equations agree. When $0<\ell_1<1$, we can replace $\ell_1 \rightarrow 1/{\ell_1}$ and reproduce \eqref{SBH-m}.}

When we integrate $*_{11}F_4$ over the seven-cycle $H$ \eqref{cycleH} associated to the Betti vector field, we have the electric baryonic charge as
\be\label{puremagnetic}
\int_H *_{11}F_4= 12N_1 -9N_2=(g-1)\left(\dfrac{L_{GKW}}{l_p} \right)^6 \dfrac{(\ell_1-\ell_2)(\ell_2+\ell_4)}{\ell_1 \ell_2^2 \ell_4\kappa} \dfrac{27}{\pi^2}.
\ee
Then one can easily see that the electric baryonic charge vanishes when $\ell_1=\ell_2$ or $\ell_2=-\ell_4$. These two cases were studied in the section \ref{dyon-BH}.
\newline

Now, let us move on to
 the second class of AdS$_2$ solutions, which are associated with non-vanishing internal four-form fluxes \eqref{flux-w4form}. In this case, the equation of the motion for the flux becomes
\be
d*_{11}F_4+\dfrac{1}{2} F_4 \wedge F_4=0.
\ee
The second term vanishes when there is the electric flux only. However, it becomes non-trivial when the magnetic flux is turned on.
Hence, we integrate $*_{11}F_4+\frac{1}{2} A_3 \wedge F_4$ over various seven-cycles and calculate the so-called Page charge, which is the correct quantity to be quantized \cite{Marolf:2000cb}.\footnote{Recently, the Page charges for rotating AdS black holes are discussed in \cite{Benini:2020gjh, Couzens:2020jgx}.
}

Let us begin with the ansatz for the three-form potential \eqref{A3ansatz} 
\be\label{A3an}
A_3=C_3- \dfrac{\sqrt{2}}{8} \left( A^1 \wedge J_{\mathbb{CP}_2} +A^2 \wedge J_{S_1^2 } \right)
+\dfrac{1}{8} \left(b_1 J_{\mathbb{CP}_2} +b_2  J_{S_1^2 }  \right) \wedge D\psi,
\ee
where
\be
D\psi=d\psi+\dfrac{1}{4} \left(P_{\mathbb{CP}_2} +P_{S_1^2 }+P_{S_2^2 }\right).
\ee
Here $P$ is the potential, which satisfies
\be
dP_{\mathbb{CP}_2}=J_{\mathbb{CP}_2},\quad dP_{S_1^2 }=J_{S_1^2 }, \quad dP_{S_2^2 }=J_{S_2^2 } .
\ee
In contrast to the magnetic black hole solutions, the dyonic black holes have non-zero $b_i$. Then, the last term in the three-form potential ansatz \eqref{A3an} gives non-trivial contributions to $\int_{M_7} A_3 \wedge F_4$ where $M_7= Y_7, \mathcal{C}_1, \mathcal{C}_2$. 
Furthermore, as explained in \cite{Donos:2014eua}, we do not have gauge potentials which are globally well-defined. Instead, we only have gauge potentials which are well-defined on each coordinate patch and related to each other via gauge transformations. Since the gauge potential \eqref{A3an} is well-defined on $Y_7$, we should find the expressions for two gauge potentials which are well-defined on $\mathcal{C}_1, \mathcal{C}_2$, respectively. Then we integrate $*_{11}F_4+\frac{1}{2} A_3 \wedge F_4$ over these seven-cycles and compute the Page charges as\footnote{In the appendix \ref{page-detail}, we calculate the Page charges for $Q^{111}$ case in detail. Here, we obtain the result for $M^{111}$ case by identifying $\ell_2=\ell_3, m_{12}=m_{13}$ and replacing the volumes of K\"ahler-Einstein manifolds appropriately. One can refer to the section 7.1.4 in \cite{Fabbri:1999hw} for the connections on the various coordinate patches in $M^{111}$.}
\begin{align}
\dfrac{1}{(2\pi l_p)^6} \int_{Y_7}G_7&=
\left(\dfrac{L_{DGK}}{l_p} \right)^6 \dfrac{\ell_2^2(\ell_1+2\ell_2)+4\ell_2 m_{12}m_{14}-\ell_1 m_{14}^2}{\ell_1 \ell_2^4}\dfrac{9}{4 \pi^2}\equiv N,\\
\dfrac{1}{(2\pi l_p)^6} \int_{\mathcal{C}_1} G_7 &=
(g-1)\left(\dfrac{L_{DGK}}{l_p} \right)^6 \dfrac{\ell_2^2(2\ell_2+\ell_4)+4 \ell_2 m_{12}m_{14}-\ell_4 m_{14}^2 }{\ell_2^4 \ell_4 \kappa}\dfrac{9}{4 \pi^2} \equiv N_1,\nn\\
\dfrac{1}{(2\pi l_p)^6} \int_{\mathcal{C}_2}G_7 &=
(g-1)\left(\dfrac{L_{DGK}}{l_p} \right)^6\nn\\& \times \dfrac{\ell_1\ell_2\ell_4(\ell_1+\ell_2+\ell_4)+(-(\ell_1-\ell_4)^2 m_{12}^2+2\ell_2(\ell_1+\ell_4) m_{12}m_{14} -\ell_2^2 m_{14}^2)}{(\ell_1 \ell_2 \ell_4)^2 \kappa}\dfrac{3}{ \pi^2},\nn\\
&\equiv N_2,\nn
\end{align}
where we define $G_7$ as
\be\label{page7}
G_7= *_{11}F_4+\dfrac{1}{2} A_3 \wedge F_4.
\ee
 Here, note that $G_7$ is not a differential form and does not define a cohomology class though it is closed, because it transforms under gauge transformations. Then, the charges obtained by integrating $G_7$ over homologous cycles are not equal. It implies that integrating $G_7$ over the trivial seven-cycle \eqref{constraint-cycle} does not become zero.
In turn, it does not reduce to the constraint \eqref{con-M111-2} imposed by the equation of motion \eqref{constraint2}.

Finally, let us calculate the magnetic fluxes as
\begin{align}
\dfrac{1}{(2\pi l_p)^4} \int_{\Sigma_g \times \mathcal{S}_1} F_4 &= \left(\dfrac{L_{DGK}}{l_p} \right)^3 \dfrac{m_{14}}{\ell_1 \ell_4}\dfrac{2}{\pi^2}\kappa,\nn\\
\dfrac{1}{(2\pi l_p)^4} \int_{\Sigma_g \times \mathcal{S}_2} F_4 &=\left(\dfrac{L_{DGK}}{l_p} \right)^3 \dfrac{m_{12}}{\ell_2 \ell_4}\dfrac{3}{\pi^2}\kappa.
\end{align}
Then the primitivity condition $2m_{12}+m_{14}=0$ enables us to define a trivial four-cycle as
\be\label{triv-four-cycle}
3 \ell_1\, \Sigma_g \times \mathcal{S}_1+4\ell_2\, \Sigma_g \times \mathcal{S}_2.
\ee
As we mentioned at the end of the section \ref{dyon-M111}, the primitivity condition of AdS$_2$ solutions in M-theory corresponds to the constraint \eqref{BH-con1}, which should be satisfied by the scalar fields of the black holes.
Certainly, one can check this correspondence also holds for the uplifted four-form flux.
Given the four-form flux \eqref{flux-uplift}, we integrate $F_4$ on this trivial cycle \eqref{triv-four-cycle} and obtain
\be
v_1(q_2-b_2 p^0 \kappa)+ 2v_2(q_1-b_1 p^0 \kappa)=0.
\ee
In addition, imposing the self-dual condition on the internal part of the four-form flux \eqref{flux-uplift}, we have
\be
(b_1+b_2) v_1^2 (q_2-b_2 p^0 \kappa)- 2b_1 v_2^2 (q_1-b_1 p^0 \kappa)=0.
\ee
Using these two equations, we easily reproduce the constraint \eqref{BH-con1}.
\subsection{Black hole entropy}
So far, we have calculated the entropies of magnetic and dyonic black holes for $M^{111}, Q^{111}$ with the presence of the four-dimensional Newton's constant. 
Now the quantization conditions of Page charges lead us to express these black hole entropies in terms of the quantized integer $N$. Let us focus on the entropy of the dyonic AdS$_4$ black holes for $Q^{111}$ \eqref{SBH-d-Q111} as 
\begin{align}
S_{\textrm{BH}}
=\dfrac{\pi}{ G_N^{(4)}} |g-1| 
\dfrac{3\sqrt{3}(\ell_1\ell_2\ell_3)^2}{16\left(\ell_1\ell_2\ell_3(\ell_1+\ell_2+\ell_3)
+\mathcal{M}\right)^{3/2}}\,\dfrac{\ell_1+\ell_2+\ell_3+\ell_4}{\ell_4 \kappa}.
\end{align}
We calculate the four-dimensional Newton's constant in \eqref{NC}
\be
\dfrac{1}{G_N^{(4)}} =\dfrac{256}{3 \sqrt{3}} \pi^2N^{3/2}
\left(\dfrac{ \ell_1 \ell_2 \ell_3(\ell_1+ \ell_2+ \ell_3)+\mathcal{M} }{\ell_1\ell_2\ell_3(\ell_1+\ell_2+\ell_3)-\mathcal{M}} \right)^{3/2} 
\left(\textrm{Vol}(S^2) \textrm{Vol(KE$_4$)} \Delta_\psi \right)^{-1/2}.
\ee
Then the final expression of black hole entropy becomes
\begin{align}\label{ent-Q111-fin}
S_{\textrm{BH}}
=|g-1| N^{3/2}
\dfrac{16\pi^3 (\ell_1\ell_2\ell_3)^2}{\left(\ell_1\ell_2\ell_3(\ell_1+\ell_2+\ell_3)
-\mathcal{M}\right)^{3/2}}\,\dfrac{\ell_1+\ell_2+\ell_3+\ell_4}{\ell_4 \kappa}
\left(\textrm{Vol}(S^2) \textrm{Vol(KE$_4$)} \Delta_\psi \right)^{-1/2}.
\end{align}
In \cite{Hong:2019wyi}, the authors already considered the Page charge, but they chose a gauge where the second term in \eqref{page7} does not contribute to the Page charge computation. As a result, they obtained the entropy depending on $\ell_i$ only. See the equation (3.22) in \cite{Hong:2019wyi}. However, we choose a gauge where the gauge potential is well-defined on $Y_7$ and obtain the non-trivial contribution from the second term in \eqref{page7}. As we can see in the equation \eqref{M-Q111}, $\mathcal{M}$ is non-trivial for the dyonic black hole. Hence, our expression for the entropy of dyonic black holes \eqref{ent-Q111-fin} does not agree with the result of \cite{Hong:2019wyi}. 
\section{Discussion}\label{discussion}
In this paper, we have revisited dyonic AdS$_4$ black holes studied in \cite{Halmagyi:2013sla} and uplifted their near horizon solutions to eleven-dimensional supergravity. Then, we identified the relations between AdS$_4$ black holes \cite{Halmagyi:2013sla} and previously known AdS$_2$ solutions in M-theory \cite{Gauntlett:2006ns, Donos:2008ug}: the uplifts of magnetic, dyonic AdS$_4$ black holes correspond to the AdS$_2$ solutions with electric, dyonic four-form fluxes, respectively. Furthermore, comparing the uplift formulae and the AdS$_2$ solutions directly, we have identified the \emph{off-shell} four-dimensional AdS black holes. More specifically, the quantities needed for describing AdS$_4$ black holes, i.e., the metric, the magnetic and electric charges and the scalars of vector- and hyper-multiplets at the horizon are determined in terms of the parameters $\ell_i, m_{ij}$ of AdS$_2$ solutions without imposing the constraint required by the equation of the motion. Hence, we have called it \emph{off-shell} AdS$_4$ black hole solution and calculated the off-shell entropy. We also have considered the flux quantizations. For the uplift solution of the dyonic black holes, AdS$_2$ solutions in M-theory has non-trivial internal four-form fluxes. Then, one should consider the quantization conditions for the Page charges, which are conserved but transform under large gauge transformations.
We have expressed the black hole entropy using this quantized integer.

As we mentioned in the introduction, AdS$_2$ solutions, which have been constructed so far, have baryonic charges. Then, it would be interesting to construct explicit AdS$_2$ solutions with nontrivial mesonic charges and compare the black hole entropy to the topologically twisted index of dual field theory. The study of new solutions with mesonic charges were tackled numerically in \cite{Hong:2019wyi}. We plan to investigate it further in the near future.

Recently, the Euclidean supergravity solutions which are called black saddles were constructed in STU supergravity \cite{Bobev:2020pjk} and their on-shell action was shown to agree exactly with the topologically twisted index of ABJM theory. It is of interest to find black saddle solutions in four-dimensional gauged supergravity constructed in \cite{Cassani:2012pj} and compare them to AdS$_4$ black holes constructed in \cite{Halmagyi:2013sla} and revisited in this paper.

Generalizing the extremization principles along the line of \cite{Couzens:2018wnk} in several directions leads to many interesting open problems.
For the dyonically charged AdS$_4$ black holes we have discussed, the first step in this program is generalizing the Gauntlett-Kim geometries \cite{Gauntlett:2007ts} to include the transgression term \cite{Donos:2008ug}.
There are also concrete examples of correspondence between the AdS$_4$ black holes in massive type IIA supergravity and dual field theories on D2-branes \cite{Guarino:2015jca, Guarino:2017eag, Guarino:2017pkw,Hosseini:2017fjo, Benini:2017oxt}. More interestingly, there have been intensive studies on the entropy of the rotating AdS black holes and their field theory computations in various dimensions. The generalization for rotating AdS$_4$ black holes was initiated in \cite{Couzens:2020jgx}. Furthermore, very recently, accelerating, rotating and dyonically charged AdS$_4$ black holes and their uplifts to $D=11$ supergravity have been studied in \cite{Ferrero:2020}.\footnote{An analogous solution in type IIB supergravity was discussed in \cite{Ferrero:2020laf}.}
It would be very nice to formulate the extremization principles for these classes of black holes.

\acknowledgments
This work was supported by the National Research Foundation of Korea (NRF) grant 
NRF-2017R1D1A1B03035515(HK), 2019R1A2C2004880(HK, NK) and 2020R1A2C1008497(HK). 
\appendix
\section{AdS$_2$ solutions in $D=11$ supergravity}\label{ads2-review}
In this section, we review AdS$_2$ solutions in $D=11$ supergravity, which were studied in \cite{Kim:2006qu} by considering M2-branes wrapped on two-cycles in a Calabi-Yau 5-fold and further developed in \cite{Gauntlett:2006ns, Donos:2008ug}. The eleven-dimensional metric ansatz is\footnote{In this section, we use $L_{GKW}$ to distinguish them from $L$ used in \eqref{metric-sol}. They are related as $L^3= L_{GKW}^3/(\ell_1+\ell_2+\ell_3+\ell_4)$. When we are dealing with the non-trivial magnetic four-form flux, $L_{GKW}$ is replaced by $L_{DGK}$.}
\be
ds_{11}^2=L_{GKW}^2 e^{2A}\biggl(ds^2(\textrm{AdS}_2)+ds^2(Y_9)\biggr).
\ee
Supersymmetry requires that the nine-dimensional internal space is a $U(1)$-fibration of eight-dimensional K\"ahler manifold.
 Here, we focus on a particular class of solution where the K\"ahler base is constructed from a product of K\"ahler-Einstein manifold $S^2$ or $H_2$. Then the metric is given by
\be
ds_{11}^2=L_{GKW}^2 e^{2A}\biggl(ds^2(\textrm{AdS}_2)+(dz+P)^2+e^{-3A}\sum_{i=1}^4 ds^2({KE_2}_{(i)})\biggr).
\ee
The two-dimensional K\"ahler-Einstein manifold is normalized such that
\be
 ds^2({KE_2}_{(i)})=\dfrac{1}{\ell_i} ds^2(\widehat{KE_2}_{(i)}),
\ee
where $\widehat{KE_2}_{(i)}$ is $T^2, S^2$ or $H_2$ with the scalar curvature $R(\widehat{KE_2}_{(i)})=2$. 
Here $dP=\mathcal{R}$ is the Ricci form for the eight-dimensional K\"ahler manifold and is given by
\be
\mathcal{R}=\sum_{i=1}^4  J_i,
\ee
where $J_i$ are K\"ahler forms of the $ds^2(\widehat{KE_2}_{(i)})$ metrics.\footnote{Our normalization is different to \cite{Gauntlett:2006ns} as $J_i^{\textrm{GKW}}=J_i^{\textrm{Ours}}/{\ell_i}$.} Note that the K\"ahler forms of eight-dimensional K\"ahler manifold is
\be
\mathcal{J}=\sum_{i=1}^4  \dfrac{1}{\ell_i}J_i.
\ee
The warp factor is given by
\be
e^{-3A}=\dfrac{R}{2}=\sum_{i=1}^4 \ell_i,
\ee
where $R$ is the Ricci scalar for eight-dimensional K\"ahler manifold. 
\subsection{Electric four-form flux}
First, let us consider AdS$_2$ solutions with the electric four-form flux as\footnote{The sign of $F_4$ is different to that of \cite{Gauntlett:2006ns}. See the equation 6.13 of that paper.}
\begin{align}
F_4&= \dfrac{L_{GKW}^3}{\ell_1+\ell_2+\ell_3+\ell_4}\textrm{vol}(AdS_2)\nn\\
&\wedge \left(
 \dfrac{\ell_2+\ell_3+\ell_4}{\ell_1}J_1+\dfrac{\ell_1+\ell_3+\ell_4}{\ell_2}J_2
 +\dfrac{\ell_1+\ell_2+\ell_4}{\ell_3}J_3+\dfrac{\ell_1+\ell_2+\ell_3}{\ell_4}\kappa J_4\right).
 \end{align}
The details can be found in the section 6.3 of \cite{Gauntlett:2006ns}.
The eight-dimensional Ricci-scalar $R$ satisfies the equations of the motion as
\be\label{theeq1}
\Box R-\dfrac{1}{2}R^2+R_{ij}R^{ij}=0.
\ee
Computing this equation, we obtain the constraint satisfied by $\ell_i$s as
\be\label{constraint1}
\sum_{i=1}^4 \ell_i^2=\left(\sum_{i=1}^4 \ell_i\right)^2,
\ee
i.e.,
\be
\ell_1 \ell_2+\ell_1 \ell_3+\ell_1 \ell_4+\ell_2 \ell_3+\ell_2 \ell_4+\ell_3 \ell_4=0.
\ee

The solutions we are interested in this paper has $Y_9$ as the $M^{111}$ fibration over $\Sigma_g$. Hence, we consider the eight-dimensional K\"ahler manifold as a product of $S^2, \mathbb{CP}^2$ and $\Sigma_g$.
Here, we set $\ell_2=\ell_3$ with $R(\mathbb{CP}^2)=4.$ Then the metic reduces to
\begin{align}\label{metric-GKW-gen}
ds_{11}^2&=L_{GKW}^2 e^{2A}
\biggl(ds^2(\textrm{AdS}_2)+(dz+P)^2\nn\\
&\phantom{=L^2 e^{2A}\biggl(}+ (\ell_1+2\ell_2+\ell_4)\biggl(\dfrac{1}{\ell_1}ds^2(S^2)+\dfrac{1}{\ell_2}ds^2(\mathbb{CP}^2) 
+ \dfrac{\kappa}{\ell_4 }ds^2(\Sigma_g)\biggr)\biggr),\\
F_4&= \dfrac{L_{GKW}^3}{\ell_1+2\ell_2+\ell_4}\textrm{vol}(AdS_2)\wedge \left(
 \dfrac{2\ell_2+\ell_4}{\ell_1}J_1+\dfrac{\ell_1+\ell_2+\ell_4}{\ell_2}J_2+\dfrac{\ell_1+2\ell_2}{\ell_4}\kappa J_4\right). \label{flux-GKW-gen}
 \end{align}
The constraint reduces to
\be\label{constraint1-M111}
2 \ell_1 \ell_2 +\ell_1 \ell_4+\ell_2^2 +2\ell_2 \ell_4=0.
\ee

By rescaling, one can set $\ell_2=1$ without loss of generality.
Solving the constraint \eqref{constraint1-M111}, we obtain 
\be
\ell_4=-\dfrac{1+2\ell_1}{2+\ell_1}.
\ee
Then the eleven-dimensional metric becomes
\begin{align}
ds_{11}^2&=L_{GKW}^2 e^{2A}
\biggl(ds^2(\textrm{AdS}_2)+(dz+P)^2\nn\\
&\phantom{=L^2 e^{2A}\biggl(}+ \dfrac{3+2\ell_1+\ell_1^2}{2+\ell_1}\biggl(\dfrac{1}{\ell_1}ds^2(S^2)+ds^2(\mathbb{CP}^2) 
+\dfrac{2+\ell_1}{1+2\ell_1}ds^2(\Sigma_g)\biggr)\biggr).
\end{align}
It is the equation (A.5) of \cite{Kim:2019umc} and (C.3) of \cite{Gauntlett:2019roi}.
\subsection{Magnetic four-form flux}\label{w4form}
Now let us turn to the case where the non-trivial internal four-form fluxes are turned on. See the section 3.2 of \cite{Donos:2008ug} for details.
The four-form flux is
\be\label{flux-w4form}
F_4 =L_{DGK}^3 \left(\textrm{vol}(AdS_2) \wedge \mathsf{F}_2 +\mathsf{F}_4 \right),
\ee
where\footnote{In the equation 3.21 of \cite{Donos:2008ug}, there is a typo in $F_2$. It should be divided by a factor 2. This typo was pointed out in \cite{Donos:2012sy}. See the footnote 4 in that paper.}
\begin{align}\label{DGK-flux}
\mathsf{F}_2&= \dfrac{1}{\sum_{i=1}^4\ell_i}\left(\dfrac{\ell_2+\ell_3+\ell_4}{\ell_1}J_1+\dfrac{\ell_1+\ell_3+\ell_4}{\ell_2}J_2+\dfrac{\ell_1+\ell_2+\ell_4}{\ell_3}J_3+\dfrac{\ell_1+\ell_2+\ell_3}{\ell_4}\kappa J_4\right),\nn\\
\mathsf{F}_4&= \sum_{i,j} \dfrac{m_{ij}}{\ell_i \ell_j} J_i \wedge J_j.
\end{align}
Here matrix $m_{ij}$ is symmetric in $i,j$ indices and their diagonal entries are zero.
When the magnetic flux is included, the equations of the motion \eqref{theeq1} is generalized to 
\be\label{theeq2}
\Box R-\dfrac{1}{2}R^2+R_{ij}R^{ij}+\dfrac{1}{4!}F_{ijkl}F^{ijkl}=0.
\ee
The equation of motion  \eqref{theeq2} reduces to the constraint satisfied by the parameters $\ell_i$ and $m_{ij}$ as
\be\label{constraint2}
\ell_1 \ell_2+\ell_1 \ell_3+\ell_1 \ell_4+\ell_2 \ell_3+\ell_2 \ell_4+\ell_3 \ell_4
=m_{12}^2+m_{13}^2+m_{14}^2+m_{34}^2+m_{24}^2+m_{23}^2.
\ee
We also have the primitive condition $\mathsf{F}_4 \wedge J=0$ which implies that
\be\label{pri-con-1}
m_{12}= m_{34},\quad m_{13}=m_{24}, \quad m_{14}=m_{23},
\ee
which is a self-dual condition, and
\be\label{pri-con-2}
m_{12}+m_{13}+m_{14}=0.
\ee
We explicitly write down the constraint with imposing a self-dual condition only as
\be\label{constraint3}
\ell_1 \ell_2+\ell_1 \ell_3+\ell_1 \ell_4+\ell_2 \ell_3+\ell_2 \ell_4+\ell_3 \ell_4
=2(m_{12}^2+m_{13}^2+m_{14}^2).
\ee

For the case KE$_4=\mathbb{CP}^2$, we identify $\ell_2=\ell_3$ and $m_{12}=m_{13}, \, m_{34}=m_{24}$. Then $\mathsf{F}_4$ becomes
\be
\mathsf{F}_4/2= \dfrac{m_{12}}{\ell_1 \ell_2} J_1 \wedge J_{\mathbb{CP}^2}
+ \dfrac{m_{14}}{\ell_1 \ell_4}\kappa J_1 \wedge J_4
+\dfrac{m_{34}}{\ell_2 \ell_4}\kappa J_{\mathbb{CP}^2} \wedge J_4
+\dfrac{m_{23}}{2\ell_2^2} J_{\mathbb{CP}^2} \wedge J_{\mathbb{CP}^2},
\ee
and the constraint reduces to
\be\label{con-M111-2}
2 \ell_1 \ell_2 +\ell_1 \ell_4+\ell_2^2 +2\ell_2 \ell_4-2(2m_{12}^2+m_{14}^2)=0.
\ee

 We set $\ell_2=1$ by rescaling and impose the remaining primitive condition. Then solving the equation of the motion \eqref{con-M111-2} gives
\be
\ell_4=-\dfrac{1+2\ell_1-2m^2}{2+\ell_1},
\ee
where $m^2\equiv 2 m_{12}^2+m_{14}^2= 6 m_{12}^2$. 
The eleven-dimensional metric becomes
\begin{align}
ds_{11}^2=&L_{DGK}^2 e^{2A}
\biggl(ds^2(\textrm{AdS}_2)+(dz+P)^2\nn\\
&+ \dfrac{3+2\ell_1+\ell_1^2+2m^2}{2+\ell_1}\biggl(\dfrac{1}{\ell_1}ds^2(S^2)+ds^2(\mathbb{CP}^2) 
- \dfrac{\kappa (2+\ell_1)}{1+2\ell_1-2m^2}ds^2(\Sigma_g)\biggr)\biggr).
\end{align}
We can easily reproduce the solutions discussed in the section \ref{dyon-BH}. 
When we have $\ell_1=1$ and introduce a new variable $w$ such that $m^2\equiv\frac{3}{2}w^2$, the metric reduces to \eqref{dyonic-kp-metric} for $\kappa=1$ case.
When we have $\ell_1=1+2m^2$ and $m^2\equiv\frac{3}{2}w^2$, the metric reduces to \eqref{dyonic-km-metric} for $\kappa=-1$ case.\footnote{The relations between these solutions and the solutions in \cite{Azzurli:2017kxo} have been studied in \cite{Hong:2019wyi}.}
\section{$Q^{111}$ solutions of Halmagyi, Petrini and Zaffaroni }\label{Q111-HPZ}
In \cite{Halmagyi:2013sla}, the authors studied the AdS black hole solutions in four-dimensional $\mathcal{N}=2$ gauged supergravity coupled to vector and hypermutiplets. As we mentioned in the main text, the prepotential $\mathcal{F}$ and the homogeneous coordinates $X^I$ are needed to describe the vector multiplet scalar manifold. 
In addition to that, one also need the Killing prepotential $P^x_{\Lambda}$ and the Killing vectors $k_\Lambda^u$ for the hypermultiplet scalar manifold. The authors considered only abelian gaugings of the hypemultiplet scalar manifold and focused on a case where
\be
P^1_{\Lambda}=P^2_{\Lambda}=0.
\ee
For the $Q^{111}$ solution we will summarize in this section, 
the Killing prepotential and the Killing vector are
\be
P_\Lambda^3= \sqrt{2} \left(4-\frac{1}{2} e^{2\phi} e_0,-e^{2\phi},-e^{2\phi},-e^{2\phi}\right),
\quad k_\Lambda^u =\sqrt{2} \left(e_0,2,2,2 \right).
\ee

Now let us explicitly record the near horizon solutions of black holes asymptotic to AdS$_4 \times Q^{111}$ studied in \cite{Halmagyi:2013sla} for readers' convenience. The solutions are parameterized by the scalars of three vector multiplets $(v_i, b_i)$, whose values are constants at the horizon, with two nontrivial constraints between them. In that sense, this solution is also off-shell. Here $i=1,\,2,\,3$. The radius of AdS$_2$ and the Riemann surface $\Sigma_g$, and a non-trivial hypermultiplet scalar $\phi$ are
\be
R_1^2= \dfrac{v_1 v_2 v_3}{16}, \quad R_2^2=\kappa R_1^2 \Big[1-\dfrac{\sigma(v_1 v_2)^2}{2\hat{\sigma}} \Big], \quad e^{2\phi}=\dfrac{4 (R_2^2 -\kappa R_1^2)}{R_2^2 \sigma(v_1 v_2)}.
\ee
The electric and magnetic charges are given by\footnote{We have corrected the typographical errors in \cite{Halmagyi:2013sla} using blue color.
} 
\begin{align}
q_0 &=\dfrac{\kappa q_{0n}}{4\sqrt{2} \hat{\sigma}},\nn\\
&\phantom{==} q_{0n}=\textcolor{blue}{-}\Big[-\sigma(v_1^3 v_3 b_{\textcolor{blue}{2}}^3)+\sigma(v_1 v_3^3 b_1^2 b_2)-(v_1 v_2 v_3)^2 \sigma(b_1)-b_1 b_2 b_3\Big(\sigma(v_1^2 b_2^2)+\sigma(v_1^2 b_2 b_3)\Big)\nn\\
&\phantom{== q_{0n}=} -v_1 v_2 v_3\Big(\sigma(v_1 b_1 b_2^2)-2\sigma(v_1 b_2^2 b_3)\textcolor{blue}{+}2\sigma(v_1^2 v_{\textcolor{blue}{3}} b_3)\Big)\Big],\nn\\
p^1 &=
\dfrac{p_n^1}{4\sqrt{2} \hat{\sigma}},\nn\\
&\phantom{==} p_n^1=2 v_1^2 v_2 v_3 (v_2^2+v_3^2+v_2 v_3)\nn\\
&\phantom{== p_n^1}+v_2 v_3(v_2^2+v_3^2)b_1^2-2 v_1 v_2 v_3 (v_2+v_3) b_2 b_3+\textcolor{blue}{1}(v_2^2+v_3^2)b_1^2 b_2 b_3+2v_1^2 b_2^2 b_3^2\nn\\
&\phantom{== p_n^1}- \Big[\Big(2v_1 v_3^2(v_2+v_3)b_1 b_{\textcolor{blue}{2}}+\Big(-v_1^2 v_2 +2 v_1 v_2 v_3 +(2v_1+v_2)v_3^2\Big)v_3 b_2^2 \Big)+ \Big(2 \leftrightarrow 3\Big) \Big]\nn\\
&\phantom{== p_n^1}+ \Big[\Big(2v_3^2 b_1 b_2^2 b_3 +(v_1^2+v_3^2)b_2^3 b_3 \Big) + \Big(2 \leftrightarrow 3\Big)\Big],\nn\\
q_1 &= \dfrac{\kappa q_{1n}}{4\sqrt{2} \hat{\sigma}},\nn\\
&\phantom{==}q_{1n}=-v_1 v_2 v_3 \sigma(v_1) b_1-\Big[v_1^2 b_2 \sigma(v_1 v_2)+ \Big(2 \leftrightarrow 3\Big) \Big]\nn\\
&\phantom{== q_{1n}} +2 v_1^2 b_1 b_2 b_3+\Big[\Big(v_2^2 b_1^3+2 v_3^2 b_1^2 b_2+(v_1^2+v_3^2) b_1 b_2^2\Big)+\Big(2 \leftrightarrow 3\Big) \Big],
\end{align}
where 
\be
\hat{\sigma}= v_1 v_2 v_3 \sigma(v_1) -\sigma(v_1^2 b_2^2)- \sigma(v_1^2 b_2 b_{\textcolor{blue}{3}}).
\ee
The polynomial $\sigma$ is defined as\footnote{Here $\sigma(1)$ implies that the index $1$ follows the permutation rules given by the elements of the symmetric group $S_3$. For example, $\sigma(v_1^2 b_2 b_3)=v_1^2 b_2 b_3+v_2^2 b_1 b_3+v_1^2 b_3 b_2+v_3^2 b_2 b_1+v_2^2 b_3 b_1+v_3^2 b_1 b_2$.}
\be
\sigma(v_1^{i_1} v_2^{i_2}v_3^{i_3}b_1^{j_1}b_2^{j_2}b_3^{j_3})=
\sum_{\sigma \in S_3}
v_{\sigma(1)}^{i_1} v_{\sigma(2)}^{i_2}v_{\sigma(3)}^{i_3}b_{\sigma(1)}^{j_1}b_{\sigma(2)}^{j_2}b_{\sigma(3)}^{j_3}.
\ee
The charges $p^2,p^3,q_2,q_3$ are similarly given by.
There are two constraints, which should be satisfied by scalar fields $(v_i, b_i)$ as
\begin{gather}
\sigma(v_1 b_2)=0,\label{BH-con1}\\
\sigma(v_1 v_2)-\sigma(b_1 b_2)=e_0\label{BH-con2}.
\end{gather}
These constraints follow from
\begin{gather}
{\cal{L}}_r^\Lambda P^3_{\Lambda}=0,\\
{\cal{L}}_i^\Lambda k_\Lambda^u =0.
\end{gather}
Here ${\cal{L}}^\Lambda$ is the symplectic section
\be
{\cal{L}}^\Lambda={\cal{L}}_r^\Lambda+ i {\cal{L}}_i^\Lambda =e^{-i \psi} e^{K/2}X^\Lambda,
\ee
where $K$ is the K\"ahler potential defined as
\be
K= -\textrm{ln}\, i \left(\bar{X}^\Lambda \mathcal{F}_\Lambda-X^\Lambda \bar{\mathcal{F}}_\Lambda \right).
\ee
The phase $\psi$ is found to be fixed to $\pi/2$.

 The Dirac quantization conditions of this class of black holes are
\be
p^\Lambda P^3_{\Lambda}= \mp 1, \quad p^\Lambda k^u_{\Lambda}=0.
\ee
Then, they give some constraints on the magnetic charges as 
\be\label{dirac-q}
p_0=\mp \dfrac{1}{4\sqrt{2}}, \quad p_1+p_2+p_3=\pm\dfrac{\sqrt{2}e_0}{16}.
\ee
Throughout this paper, we have chosen the upper sign.

The solutions for $M^{111}$ can be obtained by identifying $b_3=b_1,\, v_3=v_1$ etc.
\section{Page charges}\label{page-detail}
In this section, we provide the computational details for the quantization of the Page charges. We follow the prescription for calculating the Page charges described in \cite{Donos:2014eua}. The gauge connections cannot be globally defined. They are well-defined on each coordinate patches and related to each other via gauge transformations.

For simplicity, we consider AdS$_2$ solution in M-theory with four two-spheres, i.e. uplifting AdS$_2 \times S^2$ near horizon solution of dyonic black holes to seven-dimensional Sasaki-Einstein manifold $Q^{111}$.
First let us identify the four seven-cycles $\mathcal{C}_7^{(i)}$ obtained by fixing a point on AdS$_2$ and $S_i^2$ where $i=1,\cdots,4$. We integrate $*_{11} F_4 +\frac{1}{2} A_3 \wedge F_4$ on these cycles and compute the Page charges.

We have a globally well-defined  one-form as
\be
D\psi \equiv d\psi+\dfrac{1}{4}\left(P_1+P_2+P_3+P_4\right),
\ee
where $d P_i=J_i$. 
Let us focus on the internal part, which do not include AdS$_2$ part, of the four-form flux and the three-form potential obtained from the uplifting formula as
\begin{align}\label{F123}
F_4^{(123)}&= \dfrac{\sqrt 2}{8} \left(\alpha_{41} J_1 +\alpha_{42} J_2+ \alpha_{43} J_3 \right) \wedge J_4
+\dfrac{1}{8 } \left(\beta_{41} J_1 +\beta_{42} J_2+ \beta_{43} J_3 \right)
 \wedge  \dfrac{1}{4}\left(J_1+J_2+J_3+J_4\right),\nn\\
 A_3^{(123)}&= \dfrac{\sqrt 2}{8} \left(\alpha_{41} J_1 +\alpha_{42} J_2+ \alpha_{43} J_3 \right) \wedge P_4
+\dfrac{1}{8} \left(\beta_{41} J_1 +\beta_{42} J_2+ \beta_{43} J_3 \right)
 \wedge D\psi,
\end{align}
where 
\be
\alpha_{41}= q_1, \quad \alpha_{42}=q_2, \quad \alpha_{43}=q_3, \quad
\beta_{41}=b_1, \quad \beta_{42}=b_2, \quad \beta_{43}=b_3.
\ee
Here, we observe that the gauge potential $A_3^{(123)}$ is well-defined on $\mathcal{C}_7^{(4)}$, which is obtained by a $U(1)$-fibration over $S_1^2 \times S_2^2 \times S_3^2$. Hence we can integrate $A_3^{(123)} \wedge F_4^{(123)}$ on the seven-cycle $\mathcal{C}_7^{(4)}$ to calculate the Page charge. Now let us move on to the second seven-cycle $\mathcal{C}_7^{(3)}$. The first thing to do is finding three-form potential $A_3^{(124)}$, which is well-defined on the $\mathcal{C}_7^{(3)}$. We rewrite the four-form flux \eqref{F123}, based of the symmetries between four two-spheres, as
\begin{align}
F_4^{(124)}= \dfrac{\sqrt 2}{8} \left(\alpha_{34} J_4
 +\alpha_{31} J_1+ \alpha_{32} J_2 \right) \wedge J_3
+\dfrac{1}{8 } \left(\beta_{34} J_4 +\beta_{31} J_1+ \beta_{32} J_2 \right)
 \wedge \dfrac{1}{4} \left(J_1+J_2+J_3+J_4\right),
\end{align}
where
\begin{align}
\alpha_{34}&=\dfrac{\sqrt 2}{8}b_3-\dfrac{1}{2}\left(q_1+q_2-2q_3 \right),\quad
\alpha_{31}=\dfrac{\sqrt 2}{8}b_3-\dfrac{1}{2}\left(q_1-q_2\right),\quad
\alpha_{32}=\dfrac{\sqrt 2}{8}b_3+\dfrac{1}{2}\left(q_1-q_2\right),\nn\\
\beta_{34}&=2\sqrt{2} \left(q_1+q_2\right),\quad
\beta_{31}=b_1+2\sqrt{2} \left(q_1-q_2\right),\quad
\beta_{32}=b_2-2\sqrt{2} \left(q_1-q_2\right).
\end{align}
Then, we can easily read off the potential well-defined on $\mathcal{C}_7^{(3)}$ as
\be
 A_3^{(124)}= \dfrac{\sqrt 2}{8} \left(\alpha_{34} J_4 +\alpha_{31} J_1+ \alpha_{32} J_2 \right) \wedge P_3
+\dfrac{1}{8} \left(\beta_{34} J_4 +\beta_{31} J_1+ \beta_{32} J_2\right)
 \wedge  D\psi.
\ee
We observe that these two gauge potentials are related by a large gauge transformation
\begin{align}
A_3^{(123)}-A_3^{(134)}
&= d \Big [\dfrac{1}{8} \left(\beta_{41} P_1 +\beta_{42} P_2+\beta_{43}P_3 \right) \wedge D\psi
- \dfrac{1}{8} \left(\beta_{23} P_3 +\beta_{24} P_4+\beta_{21}P_1 \right) \wedge D\psi\nn\\
&+\dfrac{1}{8\sqrt 2} (q_3-q_1) \Big(\left(P_3-P_1 \right)\wedge \left(P_1+P_2+P_3+P_4 \right) 
-P_3 \wedge P_1\Big)\nn\\
&+\dfrac{1}{8 \sqrt 2} (q_1+q_2+q_3-3q_2)P_4 \wedge P_2\Big ].
\end{align}
Similarly, we find the three-form potential $A_3^{(134)}$ and $A_3^{(234)}$, which are well-defined on $\mathcal{C}_7^{(2)}$ and $\mathcal{C}_7^{(1)}$, respectively, as
\begin{align}
F_4^{(134)}&= \dfrac{\sqrt 2}{8} \left(\alpha_{23} J_3 +\alpha_{24} J_4+ \alpha_{21} J_1 \right) \wedge J_2
+\dfrac{1}{8 } \left(\beta_{23} J_3 +\beta_{24} J_4+ \beta_{21} J_1 \right)
 \wedge \dfrac{1}{4} \left(J_1+J_2+J_3+J_4\right),\nn\\
 A_3^{(134)}&= \dfrac{\sqrt 2}{8} \left(\alpha_{23} J_3 +\alpha_{24} J_4+ \alpha_{21} J_1  \right) \wedge P_2
+\dfrac{1}{8} \left(\beta_{23} J_3 +\beta_{24} J_4+ \beta_{21} J_1  \right)
 \wedge  D\psi,
\end{align}
where
\begin{align}
\alpha_{23}&=\dfrac{\sqrt 2}{8}b_2+\dfrac{1}{2} \left(q_1-q_3 \right),\quad
\alpha_{24}=\dfrac{\sqrt 2}{8}b_2-\dfrac{1}{2} \left(q_1-2q_2+q_3 \right),\quad
\alpha_{21}=\dfrac{\sqrt 2}{8}b_2-\dfrac{1}{2} \left(q_1-q_3 \right),\nn\\
\beta_{23}&=b_3-2\sqrt{2} \left(q_1-q_3 \right),\quad
\beta_{24}=2\sqrt{2} \left(q_1+q_3 \right),\quad
\beta_{21}=b_1+ 2\sqrt{2} \left( q_1-q_3 \right),
\end{align}
and
\begin{align}
F_4^{(234)}&= \dfrac{\sqrt 2}{8} \left(\alpha_{12} J_2 +\alpha_{13} J_3+ \alpha_{14} J_4 \right) \wedge J_1
+\dfrac{1}{8 } \left(\beta_{12} J_2 +\beta_{13} J_3+ \beta_{14} J_4 \right)
 \wedge \dfrac{1}{4} \left(J_1+J_2+J_3+J_4\right),\nn\\
 A_3^{(234)}&= \dfrac{\sqrt 2}{8} \left(\alpha_{12} J_2 +\alpha_{13} J_3+ \alpha_{14} J_4  \right) \wedge P_1
+\dfrac{1}{8} \left(\beta_{12} J_2 +\beta_{13} J_3+ \beta_{14} J_4  \right)
 \wedge  D\psi,
\end{align}
where
\begin{align}
\alpha_{12}&=\dfrac{\sqrt 2}{8}b_1-\dfrac{1}{2} \left(q_2-q_3 \right),\quad
\alpha_{13}=\dfrac{\sqrt 2}{8}b_1+\dfrac{1}{2} \left(q_2-q_3 \right),\quad
\alpha_{14}=\dfrac{\sqrt 2}{8 }b_1 +\dfrac{1}{2} \left(2q_1-q_2-q_3 \right),\nn\\
\beta_{12}&=b_2+2\sqrt{2} \left(q_2-q_3 \right),\quad
\beta_{13}=b_3-2\sqrt{2} \left(q_2-q_3 \right),\quad
\beta_{14}=2\sqrt{2} \left(q_2+q_3 \right).
\end{align}
These gauge potentials are related to $A_3^{(123)}$ by large gauge transformations as
\begin{align}
A_3^{(123)}-A_3^{(124)}
&= d \Big [\dfrac{1}{8} \left(\beta_{41} P_1 +\beta_{42} P_2+\beta_{43}P_3 \right) \wedge D\psi
- \dfrac{1}{8} \left(\beta_{34} P_4 +\beta_{31} P_1+\beta_{32}P_2 \right) \wedge D\psi\nn\\
&+\dfrac{1}{8\sqrt 2} (q_1-q_2) \Big(\left(P_1-P_2 \right)\wedge \left(P_1+P_2+P_3+P_4 \right) 
-P_1 \wedge P_2\Big)\nn\\
&+\dfrac{1}{8 \sqrt 2} (q_1+q_2+q_3-3q_3)P_4 \wedge P_3\Big ],\\
A_3^{(123)}-A_3^{(234)}
&= d \Big [\dfrac{1}{8} \left(\beta_{41} P_1 +\beta_{42} P_2+\beta_{43}P_3 \right) \wedge D\psi
- \dfrac{1}{8} \left(\beta_{12} P_2 +\beta_{13} P_3+\beta_{14}P_4 \right) \wedge D\psi\nn\\
&+\dfrac{1}{8\sqrt 2} (q_2-q_3) \Big(\left(P_2-P_3 \right)\wedge \left(P_1+P_2+P_3+P_4 \right) 
-P_2 \wedge P_3\Big)\nn\\
&+\dfrac{1}{8 \sqrt 2} (q_1+q_2+q_3-3q_1)P_4 \wedge P_1\Big ].
\end{align}

Now we compute the Page charges. First let us focus on the integration on the seven-cycle $\mathcal{C}_7^{(4)}$. 
\begin{align}
\int_{\mathcal{C}_7^{(4)}} *_{11}F_4 &=\dfrac{1}{8^3} \left(6+ 2\left(b_1 b_2+ b_2 b_3+ b_3 b_1 \right) \right) \textrm{Vol}(S_{1}^2)\textrm{Vol}(S_{2}^2)\textrm{Vol}(S_{3}^2) \Delta_{\psi},\nn\\
&= 4 L_{DGK}^6 \dfrac{\ell_1+\ell_2+\ell_3}{\ell_1 \ell_2 \ell_3} \textrm{Vol}(S_{1}^2)\textrm{Vol}(S_{2}^2)\textrm{Vol}(S_{3}^2) \Delta_{\psi},\\
\dfrac{1}{2}\int_{\mathcal{C}_7^{(4)}} A_3^{(123)} \wedge F_4
 &=\dfrac{1}{8^3} \left(2\left(b_1 b_2+ b_2 b_3+ b_3 b_1 \right) \right) \textrm{Vol}(S_{1}^2)\textrm{Vol}(S_{2}^2)\textrm{Vol}(S_{3}^2) \Delta_{\psi},\nn\\
 &=
-4 L_{DGK}^6\dfrac{1}{(\ell_1 \ell_2 \ell_3)^2}\mathcal{M}^{(4)} \textrm{Vol}(S_{1}^2)\textrm{Vol}(S_{2}^2)\textrm{Vol}(S_{3}^2) \Delta_{\psi},
 \end{align}
 where
 \begin{align}
  L_{DGK}^3&=
 \dfrac{\sqrt{3}\ell_1\ell_2 \ell_3}{32 \sqrt{\ell_1\ell_2\ell_3(\ell_1+\ell_2+\ell_3)
+\mathcal{M}}},\\
\mathcal{M}^{(4)} &\equiv \left(\ell_1^2 m_{23}^2+\ell_2^2 m_{31}^2+\ell_3^2 m_{12}^2 
 -2\ell_2 \ell_3 m_{24} m_{34}-2\ell_1 \ell_3 m_{14} m_{34}-2\ell_1 \ell_2 m_{14} m_{24}\right).\nn
 \end{align}
Here $\mathcal{M}^{(4)}$ reduces \eqref{M-Q111} when we impose the self-dual conditions on $m_{ij}$.
Then, we obtain the Page charge, which is quantized to be an integer $N_4$ as
 \begin{align}\label{page4}
\dfrac{1}{(2\pi l_p)^6}& \int_{\mathcal{C}_7^{(4)}} \left( *_{11} F_4 +\dfrac{1}{2} A_3^{(123)} \wedge F_4 \right)\nn\\
 &=\dfrac{1}{(2\pi l_p)^6}  4 L_{DGK}^6\dfrac{1}{(\ell_1 \ell_2 \ell_3)^2} \left( (\ell_1 \ell_2 \ell_3)(\ell_1+ \ell_2+ \ell_3)-\mathcal{M}^{(4)} \right)
 \textrm{Vol}(S_{1}^2)\textrm{Vol}(S_{2}^2)\textrm{Vol}(S_{3}^2) \Delta_{\psi},\nn\\
 &=\dfrac{1}{(2\pi l_p)^6}   \dfrac{3}{2^8}\dfrac{ \ell_1 \ell_2 \ell_3(\ell_1+ \ell_2+ \ell_3)-\mathcal{M}^{(4)} }{\ell_1\ell_2\ell_3(\ell_1+\ell_2+\ell_3)+\mathcal{M}}
 \textrm{Vol}(S_{1}^2)\textrm{Vol}(S_{2}^2)\textrm{Vol}(S_{3}^2) \Delta_{\psi},\nn\\
 &\equiv N_4
 \end{align}
We write down the remaining Page charges and its quantization as
\begin{align}
\dfrac{1}{(2\pi l_p)^6}& \int_{\mathcal{C}_7^{(3)}} \left( *_{11} F_4 +\dfrac{1}{2} A_3^{(124)} \wedge F_4 \right),\nn\\
 &=\dfrac{1}{(2\pi l_p)^6}  4 L_{DGK}^6\dfrac{1}{(\ell_1 \ell_2 \ell_4)^2} \left( (\ell_1 \ell_2 \ell_4)(\ell_1+ \ell_2+ \ell_4)-\mathcal{M}^{(3)} \right)
 \textrm{Vol}(S_{1}^2)\textrm{Vol}(S_{2}^2)\textrm{Vol}(S_{4}^2) \Delta_{\psi},\nn\\
 &\equiv N_3,\nn\\
 \dfrac{1}{(2\pi l_p)^6}& \int_{\mathcal{C}_7^{(2)}} \left( *_{11} F_4 +\dfrac{1}{2} A_3^{(134)} \wedge F_4 \right),\nn\\
 &=\dfrac{1}{(2\pi l_p)^6}  4 L_{DGK}^6\dfrac{1}{(\ell_1 \ell_3 \ell_4)^2} \left( (\ell_1 \ell_3 \ell_4)(\ell_1+ \ell_3+ \ell_4)-\mathcal{M}^{(2)} \right)
 \textrm{Vol}(S_{1}^2)\textrm{Vol}(S_{3}^2)\textrm{Vol}(S_{4}^2) \Delta_{\psi},\nn\\
 &\equiv N_2,\nn\\
 \dfrac{1}{(2\pi l_p)^6}& \int_{\mathcal{C}_7^{(1)}} \left( *_{11} F_4 +\dfrac{1}{2} A_3^{(234)} \wedge F_4 \right),\nn\\
 &=\dfrac{1}{(2\pi l_p)^6}  4 L_{DGK}^6\dfrac{1}{(\ell_2 \ell_3 \ell_4)^2} \left( (\ell_2 \ell_3 \ell_4)(\ell_2+ \ell_3+ \ell_4)-\mathcal{M}^{(1)} \right)
 \textrm{Vol}(S_{2}^2)\textrm{Vol}(S_{3}^2)\textrm{Vol}(S_{4}^2) \Delta_{\psi},\nn\\
 &\equiv N_1,
 \end{align}
 where we define
\begin{align}
\mathcal{M}^{(3)} &\equiv \left(\ell_1^2 m_{24}^2+\ell_2^2 m_{41}^2+\ell_4^2 m_{12}^2 
 -2\ell_1 \ell_2 m_{13} m_{23}-2\ell_1 \ell_4 m_{13} m_{43}-2\ell_2 \ell_4 m_{23} m_{43}\right),\nn\\
 \mathcal{M}^{(2)} &\equiv \left(\ell_1^2 m_{34}^2+\ell_3^2 m_{41}^2+\ell_4^2 m_{13}^2 
 -2\ell_1 \ell_3 m_{12} m_{32}-2\ell_1 \ell_4 m_{12} m_{42}-2\ell_3 \ell_4 m_{32} m_{42}\right),\nn\\
 \mathcal{M}^{(1)} &\equiv \left(\ell_2^2 m_{34}^2+\ell_3^2 m_{42}^2+\ell_4^2 m_{23}^2 
 -2\ell_2 \ell_3 m_{21} m_{31}-2\ell_2 \ell_4 m_{21} m_{41}-2\ell_3 \ell_4 m_{31} m_{41}\right).
\end{align}
We easily reproduce the results of the magnetically charged black holes by setting $m_{ij}=0$. 
For this case, we integrate the seven-form flux $ *_{11} F_4$ over the trivial seven-cycle $\mathcal{C}_7^{(1)}+\mathcal{C}_7^{(2)}+\mathcal{C}_7^{(3)}+\mathcal{C}_7^{(4)}$ and obtain
\be
 N_1+N_2+N_3+N_4=\dfrac{4}{\pi^2 l_p^6} L_{GKW}^6\dfrac{\ell_1 \ell_2+\ell_1 \ell_3+\ell_1 \ell_4+\ell_2 \ell_3+\ell_2 \ell_4+\ell_3 \ell_4}{\ell_1 \ell_2 \ell_3 \ell_4}.
\ee
It becomes zero when we recall that the integration of a closed form over a trivial cycle gives rise to zero. Then, it corresponds to the on-shell condition \eqref{constraint1}.
However, for the dyonic black hole case, the situation is quite different. As noted in \cite{Donos:2014eua}, the integrand in the definition of the Page charge is not a differential form though it is closed, because it transforms under gauge transformations. Then it does not define a cohomology class. As a result, the sum of all the Page charges is not zero and it dose not give the on-shell condition \eqref{constraint3}.
\section{Newton's constant}\label{Newton's}
In this section, we calculate the four- and two-dimensional Newton's constant.
Given the eleven-dimensional metric \eqref{metric1}, the four-dimensional Newton's constant is
\be
\dfrac{1}{G_N^{(4)}} = \dfrac{1}{G_N^{(11)}}\, \dfrac{1}{8^3}\, \textrm{Vol}(S^2)\, \textrm{Vol(KE$_4$)}\, \Delta_\psi,
\ee
where $\Delta_\psi$ is the period of $\psi$ and has the value of $\pi/2,\, \pi$ for $M^{111},\, Q^{111}$, respectively.
Here, the eleven dimensional Newton's constant is
\be
G_N^{(11)}= \dfrac{(2\pi)^8 l_p^9}{16\pi}.
\ee
For the magnetic black hole, the flux is quantized as
\be
\dfrac{1}{(2\pi l_p)^6} \int_{Y_7} * F_4= \dfrac{1}{(2\pi l_p)^6} \dfrac{3}{2^8}
\textrm{Vol}(S^2) \textrm{Vol(KE$_4$)} \Delta_\psi
\equiv N.
\ee
Using this integer $N$, the eleven-dimensional Newton's constant becomes
\be
 \dfrac{1}{G_N^{(11)}}= \dfrac{2^{17}\pi^2}{ 3\sqrt{3}}N^{3/2}
  \left(\textrm{Vol}(S^2) \textrm{Vol(KE$_4$)} \Delta_\psi \right)^{-3/2},
\ee
and, in turn, the four-dimensional Newton's constant becomes
\be\label{NC}
\dfrac{1}{G_N^{(4)}} =\dfrac{256}{3 \sqrt{3}} N^{3/2}\left(\textrm{Vol}(S^2) \textrm{Vol(KE$_4$)} \Delta_\psi \right)^{-1/2}.
\ee
In the case of the dyonic black hole, the quantity should be quantized is a Page charge as we have calculated in \eqref{page4}
\begin{align}
\dfrac{1}{(2\pi l_p)^6} &\int_{Y_7}\left( *F_4 +\dfrac{1}{2} A_3 \wedge F_4 \right)\nn\\
&=
\dfrac{1}{(2\pi l_p)^6}   \dfrac{3}{2^8}\dfrac{ \ell_1 \ell_2 \ell_3(\ell_1+ \ell_2+ \ell_3)-\mathcal{M} }{\ell_1\ell_2\ell_3(\ell_1+\ell_2+\ell_3)+\mathcal{M}}
 \textrm{Vol}(S^2) \textrm{Vol(KE$_4$)} \Delta_{\psi} 
 \equiv N.
\end{align}
As a result, the Newton's constant becomes
\be\label{NC}
\dfrac{1}{G_N^{(4)}} =\dfrac{256}{3 \sqrt{3}} \pi^2N^{3/2}
\left(\dfrac{ \ell_1 \ell_2 \ell_3(\ell_1+ \ell_2+ \ell_3)+\mathcal{M} }{\ell_1\ell_2\ell_3(\ell_1+\ell_2+\ell_3)-\mathcal{M}} \right)^{3/2} 
\left(\textrm{Vol}(S^2) \textrm{Vol(KE$_4$)} \Delta_\psi \right)^{-1/2}.
\ee

For the eleven-dimensional metric in a form of \eqref{metric2}, we calculate the black hole entropy from the two-dimensnional Newton's constant following the method described in the appendix A of \cite{Couzens:2018wnk}. The two-dimensional Newton's constant is related to the four-dimensional Newton's constant as
\begin{align}
\dfrac{1}{G_N^{(2)}}&=\dfrac{1}{G_N^{(11)}}\, \dfrac{1}{8^3}\, R_2^2\, \textrm{Vol}(\Sigma_g) \textrm{Vol}(S^2)\,  \textrm{Vol(KE$_4$)}\, \Delta_\psi,\nn\\
&=\dfrac{1}{G_N^{(4)}} R_2^2\, \textrm{Vol}(\Sigma_g).
\end{align}
Then, the Bekenstein-Hawking entropy can be written in terms of two-dimensional Newton's constant as
\be
S_{BH}=\dfrac{1}{4G_N^{(4)}} R_2^2\, \textrm{Vol}(\Sigma_g)=\dfrac{1}{4G_2}.
\ee
Given the eleven-dimensional metric \eqref{metric-GKW-gen}, we can easily compute the two-dimensional Newton's constant as
\begin{align}
\dfrac{1}{G_N^{(2)}} 
&= \dfrac{1}{G_N^{(11)}}\,L_{DGK}^9 \dfrac{\ell_1+\ell_2+\ell_3+\ell_4}{\ell_1 \ell_2 \ell_3 \ell_4}\, \textrm{Vol}(S^2)\, \textrm{Vol(KE$_4$)}\, \textrm{Vol}(\Sigma_g)\, \Delta_z,\nn\\
&=N^{3/2}
\dfrac{\ell_1+\ell_2+\ell_3+\ell_4}{ \ell_4}
\dfrac{ 32\pi^2\left(\ell_1 \ell_2 \ell_3\right)^2 }{\left(\ell_1\ell_2\ell_3(\ell_1+\ell_2+\ell_3)-\mathcal{M} \right)^{3/2}}
\left(\textrm{Vol}(S^2) \textrm{Vol(KE$_4$)} \Delta_z \right)^{-1/2}
\textrm{Vol}(\Sigma_g),
\end{align}
where $z=4 \psi$, and reproduce the Bekenstein-Hawking entropy \eqref{ent-Q111-fin}.

\bibliography{ref-uplift}{}
\end{document}